\documentclass[journal]{IEEEtran}

\input{imports}

\begin{document}

\title{Vector-based Efficient Data Hiding in Encrypted Images via Multi-MSB Replacement} 

\author{Yike Zhang,~\IEEEmembership{Student Member,~IEEE,}
        and Wenbin Luo,~\IEEEmembership{Senior Member,~IEEE}
        
\thanks{Yike Zhang and Wenbin Luo are with the Engineering Department, St. Mary's University, San Antonio,
TX, 78228, USA (e-mail: yike.zhang@vanderbilt.edu; wluo@stmarytx.edu). \emph{Source code is available upon request.}}}


\maketitle

\begin{abstract}
As an essential technique for data privacy protection, reversible data hiding in encrypted images (RDHEI) methods have drawn intensive research interest in recent years. In response to the increasing demand for protecting data privacy, novel methods that perform RDHEI are continually being developed. %
%
We propose two effective multi-MSB (most significant bit) replacement-based approaches that yield comparably high data embedding capacity, improve overall processing speed, and enhance reconstructed images' quality. Our first method, Efficient Multi-MSB Replacement-RDHEI (EMR-RDHEI), obtains higher data embedding rates (DERs, also known as payloads) and better visual quality in reconstructed images when compared with many other state-of-the-art methods. Our second method, Lossless Multi-MSB Replacement-RDHEI (LMR-RDHEI), can losslessly recover original images after an information embedding process is performed. %
%
To verify the accuracy of our methods, we compared them with other recent RDHEI techniques and performed extensive experiments using the widely accepted BOWS-2 dataset. Our experimental results showed that the DER of our EMR-RDHEI method ranged from \emrMinDER{} bit per pixel (bpp) to \emrMaxDER{} bpp with an average of \emrAverageDER{} bpp. For the LMR-RDHEI method, the average DER was \lmrAverageDER{} bpp, with a range between \lmrMinDER{} bpp and \lmrMaxDER{} bpp. %
%
Our results demonstrate that these methods outperform many other state-of-the-art RDHEI algorithms. Additionally, the multi-MSB replacement-based approach provides a clean design and efficient vectorized implementation.
\end{abstract}

\section{Introduction}
\IEEEPARstart{R}{}eversible data hiding in encrypted images (RDHEI) methods aim to protect data privacy and integrity. Digital image security plays an essential role in protecting data privacy in the military and medical industries. With the rapid development of cloud computing, more and more content owners need effective techniques to secure the data transfer process. RDHEI methods allow two separate entities to store and share information inside an image without knowing each other's identity. For this reason, RDHEI technology is urgently sought for cloud-based storage services. A cloud-based data provider (the data hider) can use RDHEI to embed information without knowing the original content. Overall, RDHEI provides both privacy and security for the content owner and the receiver.

The demand for using this technology with cloud-based storage services has made RDHEI a critical research field worldwide. Many of the current Reversible Data Hiding (RDH) methods are suitable only when the data hider is also the content owner. Despite the features of RDH, using them in a cloud-based environment may not be optimal, as the content owner and the data hider are two different entities. With the use of RDHEI, after the content owner uploads encrypted images to the cloud, the data hider can embed additional information into the encrypted images for various purposes such as tagging and inserting personal information. Over the years, researchers have developed more state-of-the-art techniques to protect data privacy, security, and integrity.

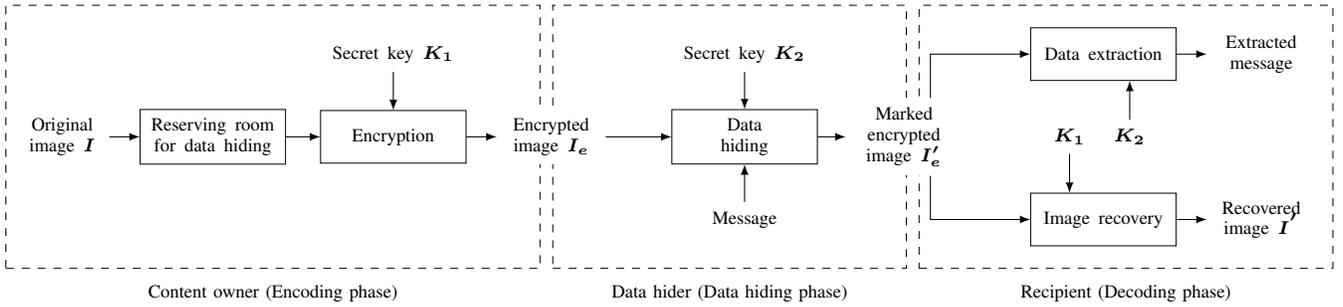
\begin{figure*}[t]
\centering

    \begin{tikzpicture}[
    dashed_box_node/.style={rectangle, draw=black, dashed, ultra thin},
    no_box_node/.style={rectangle, text width=1.2cm, align=center, minimum width=1cm, minimum height=0.2cm, fill=white},
    hard_box_node/.style={rectangle, draw=black, ultra thin, minimum width=1.7cm, minimum height=0.7cm, text width=1.7cm, align=center}]
    
    \newcommand\x{0} 
    \newcommand\y{0}
    \newcommand\boxsize{3.5cm}
    \newcommand\gridboxYshift{0.0cm}
    \newcommand\dhc{6.07} 
    \newcommand\rpc{11.34} 
    
	\node [dashed_box_node, minimum width=7.1cm, minimum height=\boxsize] at (\x, \gridboxYshift) (content_owner_d_box) {};
	\node [dashed_box_node, minimum width=4.70cm, minimum height=\boxsize] at (\x+\dhc, \gridboxYshift) (data_hider_d_box) {};
	\node [dashed_box_node, minimum width=5.50cm, minimum height=\boxsize] at (\x+\rpc, \gridboxYshift) (recipient_d_box) {};
	
	\node [below=\boxsize-1.65cm] at (\x,\y) {\scriptsize Content owner (Encoding phase)};
	\node [below=\boxsize-1.65cm] at (\x+\dhc,\y) {\scriptsize Data hider (Data hiding phase)};
	\node [below=\boxsize-1.65cm] at (\x+\rpc,\y) {\scriptsize Recipient (Decoding phase)};
	
	\node [no_box_node, text width=1cm] at (\x-2.8, \y) (original_image_n_box) {\scriptsize Original \\[-1.5mm] image \bm{$I$}};
	\node [hard_box_node, text width=1.7cm] at (\x-0.8, \y) (reserving_room_for_data_hiding_h_box) {\scriptsize Reserving room \\[-1.5mm] for data hiding};
	\node [no_box_node, text width=2cm, minimum width=2cm] at (\x+1.6, \y+1.1) (secret_key_1_n_box) {\scriptsize Secret key \bm{$K_1$}};
	\node [hard_box_node] at (\x+1.6, \y) (encryption_h_box) {\scriptsize Encryption};
	
	\node [no_box_node, minimum width = 0.8cm, text width = 1.2cm, minimum height = 0.4] at (\x+3.7, \y) (encrypted_image_n_box) {\scriptsize Encrypted \\[-1.5mm] image \bm{$I_e$}};
	
	\node [no_box_node, text width=2cm, minimum width=2cm] at (\x+\dhc+0.2, \y+1.1) (secret_key_w_n_box) {\scriptsize Secret key \bm{$K_2$}};
	\node [hard_box_node] at (\x+\dhc+0.2, \y) (data_hiding_h_box) {\scriptsize Data \\[-1.5mm] hiding};
	\node [no_box_node, text width=0.9cm] at (\x+\dhc+0.2, \y-1.1) (message_n_box) {\scriptsize Message};

	\node [no_box_node, text width=1.3cm, minimum width=0.1cm] at (\x+\rpc-2.95, \y) (marked_encrypted_image_n_box) {\scriptsize Marked \\[0mm] encrypted \\[-1.5mm] image \bm{$I_{e}'$}};
	
	\node [hard_box_node] at (\x+\rpc-0.3, \y-1.1) (image_recovery_h_box) {\scriptsize Image recovery};
	\node [hard_box_node] at (\x+\rpc-0.3, \y+1.1) (data_extraction_h_box) {\scriptsize Data extraction};
    \node [no_box_node, text width=0.5cm, minimum width=0.5cm] at (\rpc-0.75, \x) (secret_key_1_2_n_box) {\scriptsize \bm{$K_1$}};
    \node [no_box_node, text width=0.5cm, minimum width=0.5cm] at (\rpc+0.05, \x) (secret_key_2_2_n_box) {\scriptsize \bm{$K_2$}};
    \node [no_box_node] at (\x+\rpc+1.8, \y-1.1) (reconstructed_image_n_box) {\scriptsize Recovered \\[-1.5mm] image \bm{$I^{'}$}};
    \node [no_box_node] at (\x+\rpc+1.8, \y+1.1) (extracted_message_n_box) {\scriptsize Extracted \\[-1.5mm] message};
    
    \draw[arrows={-latex}] (original_image_n_box.east) -- (reserving_room_for_data_hiding_h_box.west);
    \draw[arrows={-latex}] (reserving_room_for_data_hiding_h_box.east) -- (encryption_h_box.west);
    \draw[arrows={-latex}] (secret_key_1_n_box.south) -- (encryption_h_box.north);
    
    \draw[arrows={-latex}] (encryption_h_box.east) -- (encrypted_image_n_box.west);
    \draw[arrows={-latex}] (encrypted_image_n_box.east) -- (data_hiding_h_box.west);
    
    \draw[arrows={-latex}] (secret_key_w_n_box.south) -- (data_hiding_h_box.north);
    \draw[arrows={-latex}] (message_n_box.north) -- (data_hiding_h_box.south);
    \draw[arrows={-latex}] (data_hiding_h_box.east) -- (marked_encrypted_image_n_box.west);
    
    \draw[arrows={-latex}] ([xshift=0.35cm]marked_encrypted_image_n_box.north) |- (data_extraction_h_box.west);
    \draw[arrows={-latex}] ([xshift=0.35cm]marked_encrypted_image_n_box.south) |- (image_recovery_h_box.west);
    \draw[arrows={-latex}] (secret_key_1_2_n_box.south) -- ([xshift=-0.45cm]image_recovery_h_box.north);
    \draw[arrows={-latex}] (secret_key_2_2_n_box.north) -- ([xshift=0.35cm]data_extraction_h_box.south);
    \draw[arrows={-latex}] (data_extraction_h_box.east) -- (extracted_message_n_box.west);
    \draw[arrows={-latex}] (image_recovery_h_box.east) -- (reconstructed_image_n_box.west);

\end{tikzpicture}
    
    \vspace{0.2cm}  
    
    \caption{Reserving Room Before Encryption (RRBE) Schema.}
    \label{fig:RRBE}
    
    \vspace{-0.2cm}
    
\end{figure*}

The main challenge in this research lies in finding a better trade-off between embedding capacity and maintaining the quality of the reconstructed images \citep{Puteaux2018}. A growing number of approaches have come to prominence for increasing information hiding capacity in encrypted images while preserving great visual quality during recovery. Generally, RDHEI methods can be classified into two groups: Reserving Room Before Encryption (RRBE) \citep{XZhang2016, XCao2016, KMa2013, PUTEAUX2021103085} and Vacating Room After Encryption (VRAE) \citep{JZhou2016, ZQian2016, Hong2012, PUTEAUX2021103085}. The RRBE schema is shown in Fig. \ref{fig:RRBE}. These techniques can be further divided into the following categories: prediction error \citep{SYi2018, XHu2014}, location map employment \citep{TZhang2020, Asad2020}, data compression \citep{Puteaux2020, XXie2017}, histogram modification \citep{ZYin2016, Dawen2018}, and difference expansion \citep{JChang2016, VMSekhar2019}. Moreover, some methods \citep{KMa2013, XCao2016} can perform data extraction and image decryption separately, while other approaches \citep{Hong2012, XWu2014} require simultaneous extraction and decryption.

We propose two novel RRBE-based schemas to address image recovery in both lossy and lossless RDHEI, which we have shown to outperform many state-of-the-art techniques. For our first method, Efficient Multi-MSB Replacement-RDHEI (EMR-RDHEI), the assistant information is embedded in the least significant bits (LSBs) of a given image to lower the original image's bit loss and retain the high visual quality of the reconstructed image. The other proposed method, Lossless Multi-MSB Replacement-RDHEI (LMR-RDHEI), safely inserts the assistant information into the most significant bits (MSBs) of an image. The LMR-RDHEI method recovers an image without any error; i.e., the PSNR (Peak signal-to-noise ratio) of a restored image reaches infinity. The JBIG-KIT image compression library \citep{JBIG-KIT} is used in the LMR-RDHEI method to realize lossless image recovery while achieving a comparably high data embedding capacity. Both of our multi-MSB replacement-based approaches achieve high DER and PSNR for restored images when testing with the BOWS-2 dataset.
%

    

    
    
    
    

%
In summary, the main contributions of our work are presented below:
\begin{itemize}
    \item %
    We propose two novel methods for lossy and lossless RDHEI, named EMR-RDHEI and LMR-RDHEI, respectively. The methods utilize a multi-MSB replacement-based approach, allowing large volumes of data to be embedded while retaining high/lossless reconstructed images' visual quality.
    \item %
    Compared with other state-of-the-art algorithms, the design and implementation of our methods involve lower complexity computations and do not require extra files/overhead during the transfer process, as all data is embedded directly into images.
    \item %
    We demonstrate through the benchmark BOWS-2 dataset that our methods achieve one of the highest reported DER numbers (\emrAverageDER{} bpp and \lmrAverageDER{} bpp for EMR-RDHEI and LMR-RDHEI, respectively). Moreover, the reconstructed image quality is similar to or better than many other state-of-the-art techniques.
\end{itemize}

The rest of our paper is organized as follows: Section \ref{sc:related_work} discusses other state-of-the-art approaches in this field. Section \ref{sc:methods} presents our proposed methods. Section \ref{sc:results} demonstrates experiments, security analysis, and comparison details. Finally, Section \ref{sc:conclusion} summarizes our methods' limitations and discusses future work.

\section{Related Work} \label{sc:related_work}
RDHEI is particularly useful for securing data transfer, including information authentication and protection. The RRBE pipelines, shown in Fig. \ref{fig:RRBE}, provide a general schema for the safe delivery and authentication of data. Initially, in the RRBE schema, the content owner reserves spare room in an image and encrypts it. Then, the data hider embeds secret information inside the encrypted image, and this is known as the marking process. After obtaining the marked encrypted image, the receiver decrypts the image and extracts the secret information simultaneously \citep{KMa2013, XCao2016} or separately \citep{Hong2012, XWu2014} depending on the available keys.

The first prediction-based method was proposed by \citet{Puteaux2018}, with the aim of embedding secret message in the MSB of each pixel. In their approach, predicting the MSB values without errors is a high priority. During the image reconstruction phase, they use both top and left neighboring pixels to predict the current pixel value. Considering the current pixel $p(i, j)$, with $0 \leq i \leq M-1$ and $0 \leq j \leq N-1$, for an image $P$ of size $M \times N$, the current pixel's inverse value is computed by:
\begin{equation}
    \footnotesize
    inv(i, j) = (p(i, j) + 128) \mod 256
    \label{eq:eq1}
    \normalsize
\end{equation}
The computed value $pred(i, j)$ is considered as a predictor in the final decoding phase. In order to predict the current pixel $pred(i, j)$, the authors utilized the upper pixel $p(i, j-1)$ and left pixel $p(i-1, j)$ to calculate it:
\begin{equation}
    \footnotesize
    pred(i, j) = (p(i-1, j) + p(i, j-1)) / 2
    \normalsize
\end{equation}
Next, the absolute difference between $pred(i, j)$, $p(i, j)$ and between $pred(i, j)$, $inv(i, j)$ is stored into $\Delta$ and $\Delta^{inv}$, respectively:
\begin{equation}
    \footnotesize
    \begin{cases}
        \Delta = \abs{pred(i, j) - p(i, j)} \\
        \Delta^{inv} = \abs{pred(i, j) - inv(i, j)}
    \end{cases}
    \normalsize
\end{equation}

If $\Delta \le \Delta^{inv}$, there is no prediction error since the value of $pred(i, j)$ is closer to its predictor than the inverse value $inv(i, j)$. Otherwise, there can be an error, and the corresponding information is stored in an error location binary map. 

In their paper, they applied the prediction-based technique in two different approaches: the first approach, CPE-HCRDH, yielded reconstructed images with slight distortions from the original ones, and the second approach, EPE-HCRDH, used flags to highlight incorrectly predicted pixels in order to realize lossless image recovery. According to their results, the prediction-based methods significantly increased the DER to a maximum of 1 bpp. Previously, none of the existing methods succeeded in combining high embedding capacity (near 1 bpp) and high restored images' visual quality (greater than 50 dB) \citep{Puteaux2018}. 

However, a shortcoming exists in this EPE-HCRDH method. There is a small probability (1/256) of identifying an embedded 8-bit data string ``11111111" as a flag, which will jeopardize the entire data recovery process. Our proposed schemas do not employ any fixed sentinel values as flags, and the proposed LMR-RDHEI method recovers embedded data and restores the original image perfectly during the decoding phase. 

Inspired by \citet{Puteaux2018}, \citet{Ahmad2019} presented a Prediction Error Estimation technique that leveraged how adjacent pixels in the original image are closely related and thus easier to be predicted using their neighboring pixels. In their approach, \citet{Ahmad2019} also use the upper pixel $p(i, j-1)$ and left pixel $p(i-1, j)$ to predict the current pixel $p(i, j)$. With the help of a binary location map, the available pixel locations are stored for embedding secret message's bits. Unlike \citet{Puteaux2018}'s approach, \citet{Ahmad2019} opted for Arithmetic Coding \citep{Tang2019} to compress the binary location map, embedding it into the encrypted image's LSBs. The original LSBs are stored in those MSBs that could be predicted without errors in the decoding procedure.

Their results demonstrated complete reversibility of the original image while retrieving error-free data. However, the resulting DER is still comparably low. For example, when applying the method to the grey-level image Lena of size $512 \times 512$, they only achieved the embedding rate of 0.995 bpp.

\citet{Puteaux2018}'s prediction-based method was improved by \citet{YPuyang2018} when they published a two-MSB prediction model. They introduced the Median Edge Detection (MED) predictor that generates predictions by using three neighboring pixels to acquire the prediction value $pred(i, j)$, as illustrated in the following:
\begin{equation}
    \resizebox{.91\hsize}{!}{$
    pred(i, j) =
    \begin{cases}
    \max(p(i-1, j), p(i, j-1)), p(i-1, j-1) < \min(p(i-1, j), p(i, j-1))\\
    \min(p(i-1, j), p(i, j-1)), p(i-1, j-1) > \min(p(i-1, j), p(i, j-1))\\
    p(i-1, j) + p(i, j-1) - p(i-1, j-1) \qquad\qquad\qquad\qquad\quad\; \text{otherwise}
    \end{cases}
    $}
\end{equation}


Their experimental results showed an average DER of 1.346 bpp, which is a higher embedding rate compared with \citet{Puteaux2018}'s work. However, their DER is still not optimal with the two-MSB prediction model and limits the DER to a maximum of 2 bpp.

In 2020, \citet{Puteaux2020} proposed a new recursive schema to label encrypted images by using the LOCO-I compression kit to embed bits in each bit-plane of an image. Their approach uses a grey image with size $M \times N$ pixels as a stack of 8 bit-planes $I^{[k]}$, with 0 $\leq k \leq$ 7. $I_{k}^{[k,7]}$ indicates the original images after $k$ adaptations. Then, the prediction error technique is used to predict each pixel value $p_{k}^{[k, 7]}(i, j)$ of the image $I_{k}^{[k, 7]}$, using the $7 - k$ least significant bit-planes $I_{k}^{[k+1, 7]}$ and previously scanned pixels. During this process, the first-bit value $p_{k}^{[0, 0]}$ cannot be predicted. It is kept unmodified and serves to initialize the prediction. A pixel $p_{k}^{[k, 7]}(i, j)$ from $I_{k}^{[k+1, 7]}$ is made of $8 - k$ bits, and is defined as:
\begin{equation}
    \resizebox{.5\hsize}{!}{$
    p_{k}^{[k, 7]}(i, j) = \sum_{l=k}^{7}p_{k}^{l}(i, j) \times 2^{7-l}
    $}
\end{equation} Where $p_{k}^{l}(i, j)$ is the bit at index $l$.

For their prediction-based schema, \citet{Puteaux2020} combine the Median Edge Detection predictor (also known as LOCO-I \citep{LOCO-I}) with the predictor $pred(i, j)$ of the pixel $p_{k}^{[k, 7]}(i, j)$. $pred(i,j)$ is defined in the following:

\begin{equation}
    \scriptsize
    pred(i, j) = \text{MED}(p_{k}^{[k, 7]}(i, j))\\
    = \begin{cases}
    \min{(A, B)} \quad \text{if C $\geq \max{(A, B)}$}\\
    \max{(A, B)} \quad \text{if C $\leq \min{(A, B)}$}\\
    A + B - C \quad \text{otherwise}
    \end{cases}
    \normalsize
\end{equation}
\begin{figure}[H]
\centering

    \scalebox{.7}{\begin{tikzpicture}[
    X/.style = {rectangle, draw,
                minimum width=3cm, minimum height=1cm, text width=2.8cm,
                outer sep=0cm, inner sep=0.1cm, align=center, font=\scriptsize}]

	\matrix [matrix of nodes, nodes={X, anchor=center}]
	{
        C = $p_{k}^{[k, 7]}(i-1, j-1)$ & B = $p_{k}^{[k, 7]}(i, j-1)$ \\
        A = $p_{k}^{[k, 7]}(i-1, j)$ & $p_{k}^{[k, 7]}(i, j)$ \\
	};
	
\end{tikzpicture}}
    
    \vspace{0.2cm}  
    
    \caption{Prediction of the pixel $p_{k}^{[k, 7]}(i, j)$.}
    \label{fig:PE_predictor}
    
    \vspace{-0.2cm}
    
\end{figure} 

After the predictor calculation, the inverse of $p_{k}^{[k, 7]}(i, j)$ is computed as:
\begin{equation}
    \footnotesize
    inv(i, j) = (p_{k}^{[k, 7]}(i, j) + 2^{(7-k)}) \mod{2^{8-k}}
    \normalsize
\end{equation}
$\Delta$ and $\Delta^{inv}$ are used to store the final results:
\begin{equation}
    \footnotesize
    \begin{cases}
        \Delta = \abs{pred(i, j) - p_{k}^{[k, 7]}(i, j)} \\
        \Delta^{inv} = \abs{pred(i, j) - inv(i, j)}
    \end{cases}
    \normalsize
\end{equation}

If $\Delta \leq \Delta^{inv}$, the original bit value can be predicted correctly. Otherwise, there is an error during the prediction of the current pixel --- this is noted by highlighting the PE location map $\mathbf{L}_{\mathbf{loc}}^{\mathbf{k}}$. In the paper, \citet{Puteaux2020} successfully processed all the bit-planes of an image recursively. The average payload of their recursive approach was approximately 2.4586 bpp when tested on the BOWS-2 dataset. To compare the results with our latest methods, we applied our schemas to the same benchmark dataset and achieved an average payload of \emrAverageDER{} bpp and \lmrAverageDER{} bpp for the EMR-RDHEI and LMR-RDHEI, respectively. Along with the improved data embedding rate, our methods benefit from a vectorized processing and clean design. 

\citet{HUANG2020115632} proposed an RDHEI approach based on a specific encryption process. This technique is not based on the VRAE or RRBE schema. In the encryption process, they instead compute the prediction error using the MED predictor, which is able to predict its adjacent pixel values. Assuming the original image is an 8-bit grey-level image with size $M \times N$, with $p(i, j)$ representing the original image pixels, the MED prediction process can be summarized as follows:
\begin{equation}
    \resizebox{.89\hsize}{!}{$
    pred(i, j) = 
    \begin{cases}
    \min(p(i-1, j), p(i, j-1)) &\;\;\; \text{if case 1} \\
    \max(p(i-1, j), p(i, j-1)) &\;\;\; \text{if case 2} \\
    p(i-1, j) + p(i, j-1) - p(i-1, j-1) &\;\;\; \text{else}
    \end{cases}
    $}
\end{equation} Where $2 \leq i \leq M$ and $2 \leq j \leq N$.

Additionally, case 1 and case 2 are defined as the following:
\begin{equation}
    \resizebox{.8\hsize}{!}{$
    \begin{cases}
    \text{case 1:} \quad p(i-1, j-1) \geq \max(p(i-1, j), p(i, j-1)) \\
    \text{case 2:} \quad p(i-1, j-1) \leq \min(p(i-1, j), p(i, j-1))
    \end{cases}
    $}
\end{equation}

Thus, the prediction error $e$ can be calculated as:
\begin{equation}
    \footnotesize
    e(i, j) = p(i, j) - pred(i, j)
    \normalsize
\end{equation}

Overall, their approach achieved a lossless image recovery, yet their DER is suboptimal. The average DER tested on the UCID \citep{UCID} dataset was 0.9392 bpp.

\citet{Asad2020} employed a prediction-error estimation method to vacate space before embedding a message into an image. Let $I$ be the original $M \times N$ grey-level image. They first decompose the image into four sub-images $\left(I_1, I_2, I_3, I_4\right)$ by using Eq. \ref{eq: decompose_image}.
\begin{equation}
    \footnotesize
    \begin{cases}
        I_1(i, j) = I(2i - 1, 2j - 1)\\
        I_2(i, j) = I(2i - 1, 2j)\\
        I_3(i, j) = I(2i, 2j - 1)\\
        I_4(i, j) = I(2i, 2j)\\
    \end{cases}
    \normalsize
    \label{eq: decompose_image}
\end{equation} Where $i = [1, \left \lfloor{M/2}\right \rfloor]$ and $j = [1, \left \lfloor{N/2} \right \rfloor]$.

Next, $I_1(i, j)$ is used to predict the rest of the three sub-images noted as $P_{12}(i, j)$, $P_{13}(i, j)$, and $P_{14}(i, j)$:
\begin{equation}
    \footnotesize
    P_{12}(i, j) = 
    \begin{cases}
    \ceil{(I_{1}(i, j) + I_{1}(i, j+1)) / 2} &\: \text{if $i < N / 2$} \\
    I_{1}(i, j) &\: \text{if $i = N / 2$}
    \end{cases}
    \normalsize
\end{equation}
\begin{equation}
    \footnotesize
    P_{13}(i, j) = 
    \begin{cases}
    \ceil{(I_{1}(i, j) + I_{1}(i+1, j)) / 2} &\: \text{if $i < M / 2$} \\
    I_{1}(i, j) &\: \text{if $i = M / 2$}
    \end{cases}
    \normalsize
\end{equation}
\begin{equation}
    \resizebox{.8\hsize}{!}{$
    P_{14}(i, j) = 
    \begin{cases}
    \ceil{(I_{1}(i, j) + I_{1}(i+1, j+1)) / 2} &\: \text{if $H_{1}(i, j) < H_{2}(i, j)$} \\
    \ceil{(I_{1}(i, j+1) + I_{1}(i+1, j)) / 2} &\: \text{if $H_{1}(i, j) \geq H_{2}(i, j)$} \\
    \ceil{(I_{1}(i, j) + I_{1}(i+1, j)) / 2} &\: \text{if $i < M / 2, j = N / 2$} \\
    \ceil{(I_{1}(i, j) + I_{1}(i, j+1)) / 2} &\: \text{if $i = M / 2, j < N / 2$} \\
    I_{1}(i, j) &\: \text{if $i = M / 2, j = N / 2$}
    \end{cases}
    $}
\end{equation} Then $H_1(i,j)$ and $H_2(i,j)$ are defined as: 
\begin{equation}
    \resizebox{.8\hsize}{!}{$
    \begin{cases}
        H_{1}(i, j) = \abs{I_{1}(i, j) - I_{1}(i+1, j+1)} \quad (i < M / 2, j < N / 2) \\
        H_{2}(i, j) = \abs{I_{1}(i, j+1) - I_{1}(i+1, j)} \quad (i < M / 2, j < N / 2)
    \end{cases}
    $}
\end{equation}
Afterwards, the prediction-error values of the sub-images are estimated as:
\begin{equation}
    \footnotesize
    \begin{cases}
       PE_{1}(i, j) = I_{1}(i, j) \\
       PE_{2}(i, j) = I_{2}(i, j) - PE_{12}(i, j) &\quad i = 1, 2, ..., M/2\\
       PE_{3}(i, j) = I_{3}(i, j) - PE_{13}(i, j) &\quad j = 1, 2, ..., N/2 \\
       PE_{4}(i, j) = I_{4}(i, j) - PE_{14}(i, j)\\
    \end{cases}
    \normalsize
\end{equation}
After that, a location map is created to store prediction-error values' places to determine whether a particular pixel can be safely used for data embedding. 

In \citet{Asad2020}'s approach, they are able to recover an image losslessly. However, the DER of their approach was less than 0.75 bpp.

Our proposed schema employs a location map as an indicator to help us track redundancy in a given image. However, our location map generation process is performed in a completely different manner (via multi-MSB replacement). With our proposed location map, our methods achieve significantly higher embedding rates and high visual quality in restored images.

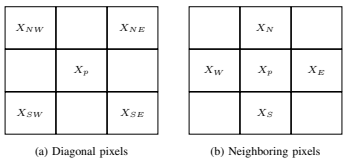
\begin{figure}[H]

    \captionsetup[subfigure]{justification=centering}
    \centering
    
    \begin{adjustbox}{minipage=\linewidth,scale=0.56}
    \subfloat[Diagonal pixels]{
        \label{fig:wcbp_a}
        \centering
        \begin{tikzpicture}[
    X/.style = {rectangle, draw,
                minimum width=1.2cm, minimum height=1cm, text width=1cm,
                outer sep=0cm, inner sep=0.1cm, align=center, font=\scriptsize}]

	\matrix [matrix of nodes, nodes in empty cells, nodes={X, anchor=center}]
	{
        $X_{NW}$ & & $X_{NE}$\\
         & $X_{p}$ & \\
        $X_{SW}$ & & $X_{SE}$\\
	};
	
\end{tikzpicture}
    }
    \subfloat[Neighboring pixels]{
        \label{fig:wcbp_b}
        \centering
        \begin{tikzpicture}[
    X/.style = {rectangle, draw,
                minimum width=1.2cm, minimum height=1cm, text width=1cm,
                outer sep=0cm, inner sep=0.1cm, align=center, font=\scriptsize}]

	\matrix [matrix of nodes, nodes in empty cells, nodes={X, anchor=center}]
	{
         & $X_{N}$ & \\
        $X_{W}$ & $X_{p}$ & $X_{E}$\\
         & $X_{S}$ & \\
	};
	
\end{tikzpicture}
    }

    \caption{WCBP examples.}
    \label{fig:wcbp}
    \end{adjustbox}
    
    \vspace{-0.2cm}
    
\end{figure}

\citet{SYi2018} proposed a method based on prediction-error encoding (PE-RDHEI) that uses a weighted checkerboard prediction (WCBP) to predict $3/4$ of the pixels in an original image with the help of the remaining $1/4$ of the pixels.
Given an 8-bit $M \times N$ grey-level image $I$, they first separate the pixels into two categories $I_1$ and $I_2$. $I_1$ consists of the $MN/4$ pixels and $I_2$ contains the remaining $3MN/4$ pixels. The first step is to predict the pixels (Eq. \ref{eq: wcbp_a}) based on its four diagonal pixels (Fig. \ref{fig:wcbp}\subref{fig:wcbp_a}). The second step is to predict the pixels (Eq. \ref{eq: wcbp_b}) by its four neighboring pixels (Fig. \ref{fig:wcbp}\subref{fig:wcbp_b}):
\begin{equation}
    \resizebox{.89\hsize}{!}{$
    X_{p} = 
    \begin{cases}
       \nint{0.35 \times (X_{NW} + X_{SE}) + 0.15 \times (X_{NE} + X_{SW})} &\;\;\; \text{if }\abs{X_{NW} - X_{SE}} < \abs{X_{NE} - X_{SW}} \\ 
       \nint{0.15 \times (X_{NW} + X_{SE}) + 0.35 \times (X_{NE} + X_{SW})} &\;\;\; \text{otherwise}
    \end{cases}
    $}
    \label{eq: wcbp_a}
\end{equation}
\begin{equation}
    \resizebox{.85\hsize}{!}{$
    X_{p} = 
    \begin{cases}
       \nint{0.35 \times (X_{N} + X_{S}) + 0.15 \times (X_{W} + X_{E})} &\;\;\; \text{if }\abs{X_{N} - X_{S}} < \abs{X_{W} - X_{E}} \\ 
       \nint{0.15 \times (X_{N} + X_{S}) + 0.35 \times (X_{W} + X_{E})} &\;\;\; \text{otherwise}
    \end{cases}
    $}  
    \label{eq: wcbp_b}
\end{equation}

Using the WCBP method, they realized a lossless image recovery with an average DER of 1.907 bpp. We compared our LMR-RDHEI with their method and obtained an average DER of \lmrAverageDER{} when testing on a large dataset of 10,000 images.

In 2019, \citet{Liu2019} published a fully reversible RDHEI schema based on pixel-prediction techniques. They use gradient-adjust predictions to detect spare space in the original image and hide secret information inside it. Given the original image $I$ of size $M \times N$, the predicted pixel value $p^{'}(i,j)$ can be obtained from the original pixels $p(i,j)$, as explained in the following:

\begin{equation}
    \resizebox{.89\hsize}{!}{$
    p^{'}(i, j) = 
    \begin{cases}
       p(i, j-1) &\;\;\; \text{if } d_{v} - d_{h} > 80 \\ 
       p(i-1, j) &\;\;\; \text{if } d_{v} - d_{h} < -80 \\ 
       (p(i-1, j)+p(i, j-1))/2 + \\ (p(i-1, j+1) + p(i-1, j-1))/4 &\;\;\; \text{otherwise}
    \end{cases},
    $} 
\end{equation} where $d_{v}$ and $d_{h}$ denotes the gradient change in the vertical and horizontal directions. They are defined in Eq. \ref{eq: Liu2019}.
\begin{equation}
     \resizebox{.89\hsize}{!}{$
    \begin{cases}
    d_{v} = \abs{p(i, j-1) - p(i, j-2)} + \abs{p(i-1, j) - p(i-1, j-1)} + \\ \;\qquad \abs{p(i-1, j) - p(i-1, j+1)} \\
    d_{h} = \abs{p(i, j-1) - p(i-1, j-1)} + \abs{p(i-1, j) - p(i-2, j)} + \\ \;\qquad \abs{p(i-1, j+1) - p(i-2, j+1)}
    \end{cases},
    $}
    \label{eq: Liu2019}
\end{equation} where $3 \leq i \leq M$ and $3 \leq j \leq N-1$.\footnote{This is due to the prediction limitations as there are not enough surrounding pixels.} The final results showed that their method could achieve an average DER of 0.95 bpp. One of the merits of their schema is realizing a lossless recovery among restored images.



Most recently, \citet{Wang2021} proposed an RDHEI method based on adaptive MSB prediction. The original image is first encrypted in a block-wise manner, so the pixels' correlation within the block can be preserved. In their prediction, the targeted image is divided into $2 \times 2$ non-overlapping blocks. Finally, they achieved a lossless original image recovery at an average DER of 2.26 bpp when testing on a selected set of 100 grey-level images from the BOWS-2 dataset.

In summary, the recently published methods discussed above have common shortcomings, such as sacrificing lower DER for higher reconstructed images' visual quality and vice versa. Our proposed methods achieved both higher DER and higher recovered images' visual quality, as illustrated in Section \ref{sc:methods} and evaluated in Section \ref{sc:results}.

\section{Proposed Multi-MSB Replacement Methods} \label{sc:methods}
\begin{figure*}[t]
\centering

    \begin{tikzpicture}[
    dashed_box_node/.style={rectangle, draw=black, dashed, ultra thin},
    no_box_node/.style={rectangle, text width=1.2cm, align=center, minimum width=1cm, minimum height=0.2cm, fill=white},
    hard_box_node/.style={rectangle, draw=black, ultra thin, minimum width=1.7cm, minimum height=0.7cm, text width=1.7cm, align=center}]
    
    \newcommand\x{0} 
    \newcommand\y{0.5}
    \newcommand\boxsize{4.1cm}
    \newcommand\gridboxYshift{0.0cm}
    \newcommand\dhc{6.07} 
    \newcommand\rpc{11.34} 
    
	\node [dashed_box_node, minimum width=7.1cm, minimum height=\boxsize] at (\x, \gridboxYshift) (content_owner_d_box) {};
	\node [dashed_box_node, minimum width=4.70cm, minimum height=\boxsize] at (\x+\dhc, \gridboxYshift) (data_hider_d_box) {};
	\node [dashed_box_node, minimum width=5.50cm, minimum height=\boxsize] at (\x+\rpc, \gridboxYshift) (recipient_d_box) {};
	
	\node [below=\boxsize-1.3cm] at (\x,\y) {\scriptsize Encoding phase};
	\node [below=\boxsize-1.3cm] at (\x+\dhc,\y) {\scriptsize Data hiding phase};
	\node [below=\boxsize-1.3cm] at (\x+\rpc,\y) {\scriptsize Decoding phase};
	
	\node [no_box_node, text width=1cm] at (\x-2.8, \y) (ORIGINAL_IMAGE) {\scriptsize Original \\[-1.5mm] image \bm{$I$}};
	\node [hard_box_node, text width=1.7cm] at (\x-1, \y) (I_ROTATION) {\scriptsize $I$ rotation};
	\node [no_box_node, text width=2cm, minimum width=2cm] at (\x-1, \y+1.1) (SECRET_KEY_1_0) {\scriptsize Secret key \bm{$K_1$}};
	\node [hard_box_node, text width=1.7cm] at (\x-1, \y-1.2) (LOCATION_MAP_GENERATION) {\scriptsize \scriptsize Location map \\[-1.5mm] generation};
	\node [no_box_node, text width=2cm, minimum width=2cm] at (\x+1.6, \y+1.1) (SECRET_KEY_1_1) {\scriptsize Secret key \bm{$K_1$}};
	\node [hard_box_node] at (\x+1.6, \y) (encryption_h_box) {\scriptsize Encryption};
	
	\node [no_box_node, text width=1cm, minimum width=1cm, minimum width=0.1cm, minimum height=0.1cm,] at (\x+2.25, \y-0.6) (embedding) {\scriptsize embedding};
	
	\node [hard_box_node] at (\x+1.6, \y-1.2) (LOCATION_MAP_ROTATION) {\scriptsize Location map \\[-1.5mm] rotation};
	\node [no_box_node, text width=2cm, minimum width=2cm] at (\x+1.6, \y-2.3) (SECRET_KEY_1_2) {\scriptsize Secret key \bm{$K_1$}};
	
	\node [no_box_node, minimum width = 0.8cm, text width = 1.2cm, minimum height = 0.4] at (\x+3.7, \y) (encrypted_image_n_box) {\scriptsize Encrypted \\[-1.5mm] image \bm{$I_e$}};
	
	\node [no_box_node, text width=2cm, minimum width=2cm] at (\x+\dhc+0.2, \y+1.1) (secret_key_w_n_box) {\scriptsize Secret key \bm{$K_2$}};
	\node [hard_box_node] at (\x+\dhc+0.2, \y) (data_hiding_h_box) {\scriptsize Data \\[-1.5mm] hiding};
	\node [no_box_node, text width=0.9cm] at (\x+\dhc+0.2, \y-1.8) (message_n_box) {\scriptsize Message};

	\node [no_box_node, text width=1.3cm, minimum width=0.1cm] at (\x+\rpc-2.95, \y) (marked_encrypted_image_n_box) {\scriptsize Marked \\[0mm] encrypted \\[-1.5mm] image \bm{$I_{e}'$}};
	
	\node [hard_box_node] at (\x+\rpc-0.3, \y-1.8) (image_recovery_h_box) {\scriptsize Image recovery};
	\node [hard_box_node] at (\x+\rpc-0.3, \y+0.8) (data_extraction_h_box) {\scriptsize Data extraction};
    \node [no_box_node, text width=0.5cm, minimum width=0.5cm] at (\rpc-0.75, \x) (secret_key_1_2_n_box) {\scriptsize \bm{$K_1$}};
    \node [no_box_node, text width=0.5cm, minimum width=0.5cm] at (\rpc+0.05, \x) (secret_key_2_2_n_box) {\scriptsize \bm{$K_2$}};
    \node [no_box_node] at (\x+\rpc+1.8, \y-1.8) (reconstructed_image_n_box) {\scriptsize Recovered \\[-1.5mm] image \bm{$I^{'}$}};
    \node [no_box_node] at (\x+\rpc+1.8, \y+0.8) (extracted_message_n_box) {\scriptsize Extracted \\[-1.5mm] message};

    \draw[arrows={-latex}] ([xshift=-0.1cm]ORIGINAL_IMAGE.east) -- (I_ROTATION.west);
    \draw[arrows={-latex}] (I_ROTATION.east) -- (encryption_h_box.west);
    \draw[arrows={-latex}] (LOCATION_MAP_ROTATION.north) -- (encryption_h_box.south);
    \draw[arrows={-latex}] (ORIGINAL_IMAGE.south) |- (LOCATION_MAP_GENERATION.west);
    \draw[arrows={-latex}] (LOCATION_MAP_GENERATION.east) -- (LOCATION_MAP_ROTATION.west);
    \draw[arrows={-latex}] (SECRET_KEY_1_0.south) -- (I_ROTATION.north);
    \draw[arrows={-latex}] (SECRET_KEY_1_1.south) -- (encryption_h_box.north);
     \draw[arrows={-latex}] (SECRET_KEY_1_2.north) -- (LOCATION_MAP_ROTATION.south);
    
    \draw[arrows={-latex}] (encryption_h_box.east) -- (encrypted_image_n_box.west);
    \draw[arrows={-latex}] (encrypted_image_n_box.east) -- (data_hiding_h_box.west);
    
    \draw[arrows={-latex}] (secret_key_w_n_box.south) -- (data_hiding_h_box.north);
    \draw[arrows={-latex}] (message_n_box.north) -- (data_hiding_h_box.south);
    \draw[arrows={-latex}] (data_hiding_h_box.east) -- ([xshift=0.15cm]marked_encrypted_image_n_box.west);
    
    \draw[arrows={-latex}] ([xshift=0.35cm]marked_encrypted_image_n_box.north) |- (data_extraction_h_box.west);
    \draw[arrows={-latex}] ([xshift=0.35cm]marked_encrypted_image_n_box.south) |- (image_recovery_h_box.west);
    \draw[arrows={-latex}] (secret_key_1_2_n_box.south) -- ([xshift=-0.45cm]image_recovery_h_box.north);
    \draw[arrows={-latex}] (secret_key_2_2_n_box.north) -- ([xshift=0.35cm]data_extraction_h_box.south);
    \draw[arrows={-latex}] (data_extraction_h_box.east) -- ([xshift=0.15cm]extracted_message_n_box.west);
    \draw[arrows={-latex}] (image_recovery_h_box.east) -- ([xshift=0.15cm]reconstructed_image_n_box.west);

\end{tikzpicture}
    
    \vspace{0.2cm}  
    
    \caption{General Schema of EMR-RDHEI Method.}
    \label{fig:General_EMR}
    
    \vspace{-0.2cm}
    
\end{figure*}
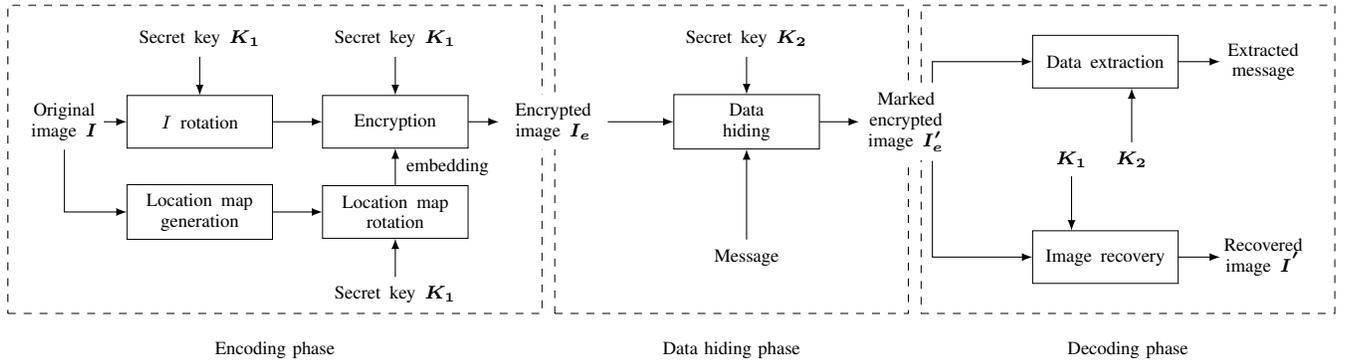
In our paper, we present two methods: EMR-RDHEI and LMR-RDHEI. With EMR-RDHEI, we are able to achieve high DER and high visual quality in restored images. With our LMR-RDHEI method, images are restored perfectly without any loss. The proposed multi-MSB replacement methods differ from the prediction-based approach employed by \citet{Puteaux2018, Puteaux2020} and \citet{YPuyang2018}. 

We employ a one-to-one binary location map that tracks redundant information in the original image. Each pixel in the input image is assigned a label of 0 or 1 in the associated binary location map depending on the multiple MSBs comparison. Instead of performing expensive numerical operations to predict the MSB value \citep{Puteaux2018, YPuyang2018, Puteaux2020}, we directly compare the MSBs of adjacent pixels to determine if they are identical and thereby consider them as redundant. By detecting and using these redundancies within the images' MSBs, we can embed large amounts of information with near-zero or zero loss of original data --- depending on the method we apply.

In the proposed methods, we directly embed our assistant information into an encrypted image. In EMR, the assistant information is the location map. In LMR, it contains both the location map and the first MSB map\footnote{This is extracted from the first MSB bit-plane of an image.}. Once the assistant information is generated, it may contain partial textures from the original image. Thus, we apply the rotation procedure to the original image and assistant information to break potential patterns. This process further enhances the overall data security. More details are presented in Section \ref{ssc:detailed_examples}.

Since modern computers can perform bit comparison operations much faster than other numerical computations, e.g., division and multiplication, our algorithms can be executed more efficiently than many other state-of-the-art methods. The number, $b$, of MSBs ($b$-MSBs) used in our methods is heavily dependent on an image's texture. In our proposals, two or more MSBs ($b \geq 2$) of a pixel are used for storing data through redundancy detection when considering adjacent pixels. This way, our methods achieve high embedding capacity and low/no data loss.

\subsection{EMR-RDHEI method}
The general schema of our EMR-RDHEI method, shown in Fig. \ref{fig:General_EMR}, includes three phases: encoding, data hiding, and decoding. The encoding phase in EMR-RDHEI is composed of five steps: location map generation, location map rotation, original image rotation, original image encryption, and location map embedding. The rotation and encryption sub-processes use the secret key $K_1$ to rotate the original image/location map and encrypt the original image. In the data hiding phase, we employ the multi-MSB replacement technique to embed a secret message into the encrypted image's redundant pixel bits. After obtaining the marked encrypted image (encrypted image with the secret message) in the decoding phase, the hidden message is extracted and recovered without any errors, and the original image is reconstructed. Moreover, the EMR-RDHEI method is separable, meaning that we can perform the data extraction and image reconstruction independently. During all phases, the images are processed in a vectorized sliding window manner. 

\subsubsection{Encoding phase}

\paragraph{Optimal location map generation} \label{ph:optimal_location_map_generation}
An essential part of our proposed methods is generating an optimal location map for an original image $I$ that finds the best trade-off between parameter $b$ and redundant pixels quantity. Before embedding a secret message into an image, we identify the redundant pixels whose multiple MSBs ($b$-MSBs) can be replaced and used for data embedding. 

\begin{figure}[H]
\centering
    \scalebox{.55}{\input{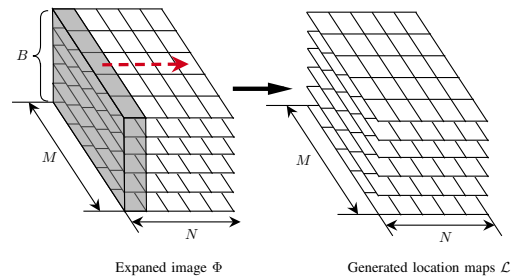}}
    
    \vspace{0.2cm}  
    
    \caption{Location map generation process.}
    \label{fig: expaned_image}
    
    \vspace{-0.2cm}
    
\end{figure}

As shown in Fig. \ref{fig: expaned_image}, the original image $I$ of size $M \times N$ is expanded via a new dimension (noted as B), resulting in the expanded image $\Phi$ of size ($M \times N \times B$). This dimensional expansion allows the parallel comparison of multiple $b$ parameters by gliding matrix $\Omega$ ($B \times M$) along the $N$ dimension. $\Omega$, a matrix that tracks the last non-redundant MSBs, is used to identify changes in MSBs from pixel to pixel. The detected redundancies via the MSBs comparisons are then stored in the location maps $\mathcal{L}$ for all possible $b$ parameters ($b \in [2, 7], b \in \mathbf{Z}$). Once the entire image is processed, the DER of each generated location map is individually calculated; thus, we can select the location map with the highest DER for the input image.
\\

At the beginning of the generation process, $\Omega$ is initialized based on the first column of $\Phi$, and the first column of $\mathcal{L}$ is set with all 1s,  defined as follows:
\begin{align}
    &\Omega(i, k) = \Phi(i, 0, k) \\
    &\mathcal{L}(i, 0, k) = 1
\end{align}

As $\Omega$ glides along the $N$ dimension, $\mathcal{L}$ is generated based on the values of $\Phi$ and $\Omega$. If their $[1,k+1]$ MSBs are equal, then $\mathcal{L} (i, j, k)$ is marked as 0 (redundant); otherwise, $\mathcal{L} (i, j, k)$ is marked as 1 (non-redundant). After each $\Omega$ and $\Phi$ comparison, $\Omega$ is consistently updated based on the newest non-redundant MSBs. Eq. \ref{eq: gliding1} and Eq. \ref{eq: gliding2} elaborate on the updating operations executed while gliding $\Omega$ along the $N$ dimension. 
\begin{equation}
    \resizebox{.88\hsize}{!}{$
    \mathcal{L}(i,j,k) = \mathbb{I}[\Phi^{[1, k+1]}(i,j,k) = \Omega^{[1, k+1]}(i,k)], j \in [1, N-1], j \in \mathbf{Z}
    $}
    \label{eq: gliding1}
\end{equation}
\begin{equation}
    \resizebox{.88\hsize}{!}{$
    \Omega(i,k) := \begin{cases}
    \Phi(i,j,k) & \text{if} \;\; \mathbb{I}[\mathcal{L}(i,j,k) = 1], j \in [1, N-1], j \in \mathbf{Z}\\
    \Omega(i,k) & \text{otherwise}
    \end{cases}
    $}
    \label{eq: gliding2}
\end{equation}Where $\mathbb{I}$ is the indicator function; $i$, $j$, and $k$ are the indices of the tensor $M$, $N$, and $B$, respectively.

To find the most optimal location map, we calculate the DER of each location map stacked in $\mathcal{L}$. We select the location map along with the parameter $b$ with the highest DER. The process of selecting the optimal $b$ is interpreted as Eq. \ref{eq: optimial_b}:
\begin{equation}
    \resizebox{.88\hsize}{!}{$
    b = [\argmax\limits_{k} \frac{1}{M \times N} \sum\limits_{i = 0}^{M-1}\sum\limits_{j = 0}^{N-1}\mathbb{I}(\mathcal{L}(i,j,k) = 0) \times b] + 2
    $}
    \label{eq: optimial_b}
\end{equation} Where $k \in [0, 5]$, $b \in [2, 7]$, $k \in \mathbf{Z}$, $b \in \mathbf{Z}$, and $b = k+2$.

\begin{figure}[H]
\centering
    \scalebox{.9}{\input{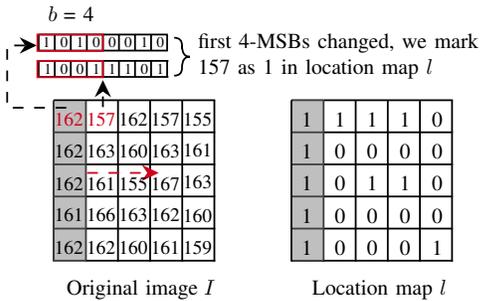}}
    
    \vspace{0.2cm}  
    
    \caption{Location map $l$ generation ($b$ = 4).}
    \label{fig: location_map}
    
    \vspace{-0.2cm}
    
\end{figure}

To further elaborate on the process of generating a location map, we go through the procedure of constructing only one location map $l$ with $b$ set to $4$ (shown in Fig. \ref{fig: location_map}). As mentioned in the initialization process, the first column of $l$ is set to all 1s, and $\omega$ is set to the first column of $I$. As the $\omega$ vector (red arrow in Fig. \ref{fig: location_map}) glides along the image width $N$, comparisons with the $b$-MSBs of $\omega$ and $I$ are performed. If redundancy is detected, i.e., no change in MSBs, we mark that pixel as a 0 in $l$; otherwise, we mark it as a 1. When the location map is complete, the image's redundancy is used to embed a secret message by replacing their tracked redundant MSBs. 

\paragraph{Secret key generation} \label{ph:secret_key_generation}
After obtaining the most optimal location map, the content owner needs to select or generate an encryption key ($K_1$) to encrypt their original image. Either 2D-LSCM (2D Logistic-Sine-coupling map \citep{2DLSCM}) function or chaotic generator (based on the Piecewise Linear Chaotic Map \citep{chaotic}) can be used to generate a sequence of pseudo-random bytes. By using the keystream generator, a long sequence of generated bytes $s(i, j)$ is constructed and reshaped into the input image's size. In our proposed encryption schema, the only requirement is to use a cryptographically secure stream cipher while encrypting. This high-quality cipher can be acquired not only by using the methods mentioned above but through RC4 and its variants suggested in \cite{Spritz2016, Souradyuti2004, Bartosz2004, Subhamoy2008}.

\begin{figure}[ht]
\centering

    \scalebox{.5}{\newcommand{\SCALE}{0.95}

\tikzset{every picture/.style={line width=0.75pt}} 

\begin{tikzpicture}[x=0.75pt,y=0.75pt,yscale=-\SCALE,xscale=\SCALE,every node/.style={scale=1*\SCALE}]

\draw  [draw opacity=0] (56,142) -- (217,142) -- (217,302.5) -- (56,302.5) -- cycle ; \draw   (56,142) -- (56,302.5)(76,142) -- (76,302.5)(96,142) -- (96,302.5)(116,142) -- (116,302.5)(136,142) -- (136,302.5)(156,142) -- (156,302.5)(176,142) -- (176,302.5)(196,142) -- (196,302.5)(216,142) -- (216,302.5) ; \draw   (56,142) -- (217,142)(56,162) -- (217,162)(56,182) -- (217,182)(56,202) -- (217,202)(56,222) -- (217,222)(56,242) -- (217,242)(56,262) -- (217,262)(56,282) -- (217,282)(56,302) -- (217,302) ; \draw    ;
\draw  [draw opacity=0][fill={rgb, 255:red, 255; green, 255; blue, 255 }  ,fill opacity=1 ] (486,142) -- (647,142) -- (647,302.5) -- (486,302.5) -- cycle ; \draw   (486,142) -- (486,302.5)(506,142) -- (506,302.5)(526,142) -- (526,302.5)(546,142) -- (546,302.5)(566,142) -- (566,302.5)(586,142) -- (586,302.5)(606,142) -- (606,302.5)(626,142) -- (626,302.5)(646,142) -- (646,302.5) ; \draw   (486,142) -- (647,142)(486,162) -- (647,162)(486,182) -- (647,182)(486,202) -- (647,202)(486,222) -- (647,222)(486,242) -- (647,242)(486,262) -- (647,262)(486,282) -- (647,282)(486,302) -- (647,302) ; \draw    ;
\draw  [color={rgb, 255:red, 0; green, 0; blue, 0 }  ,draw opacity=1 ][fill={rgb, 255:red, 155; green, 155; blue, 155 }  ,fill opacity=0.5 ] (56,142) -- (96,142) -- (96,182) -- (56,182) -- cycle ;
\draw  [fill={rgb, 255:red, 74; green, 74; blue, 74 }  ,fill opacity=0.65 ] (136,182) -- (176,182) -- (176,222) -- (136,222) -- cycle ;
\draw  [color={rgb, 255:red, 0; green, 0; blue, 0 }  ,draw opacity=1 ][fill={rgb, 255:red, 74; green, 74; blue, 74 }  ,fill opacity=0.65 ] (96,142) -- (136,142) -- (136,182) -- (96,182) -- cycle ;
\draw  [fill={rgb, 255:red, 155; green, 155; blue, 155 }  ,fill opacity=0.5 ] (176,262) -- (216,262) -- (216,302) -- (176,302) -- cycle ;
\draw  [fill={rgb, 255:red, 155; green, 155; blue, 155 }  ,fill opacity=0.5 ] (56,222) -- (96,222) -- (96,262) -- (56,262) -- cycle ;
\draw  [fill={rgb, 255:red, 155; green, 155; blue, 155 }  ,fill opacity=0.5 ] (176,182) -- (216,182) -- (216,222) -- (176,222) -- cycle ;
\draw  [fill={rgb, 255:red, 74; green, 74; blue, 74 }  ,fill opacity=0.65 ] (96,222) -- (136,222) -- (136,262) -- (96,262) -- cycle ;
\draw  [fill={rgb, 255:red, 74; green, 74; blue, 74 }  ,fill opacity=0.65 ] (56,182) -- (96,182) -- (96,222) -- (56,222) -- cycle ;
\draw  [fill={rgb, 255:red, 74; green, 74; blue, 74 }  ,fill opacity=0.65 ] (176,142) -- (216,142) -- (216,182) -- (176,182) -- cycle ;
\draw  [color={rgb, 255:red, 0; green, 0; blue, 0 }  ,draw opacity=1 ][fill={rgb, 255:red, 155; green, 155; blue, 155 }  ,fill opacity=0.5 ] (96,182) -- (136,182) -- (136,222) -- (96,222) -- cycle ;
\draw  [color={rgb, 255:red, 0; green, 0; blue, 0 }  ,draw opacity=1 ][fill={rgb, 255:red, 155; green, 155; blue, 155 }  ,fill opacity=0.5 ] (136,142) -- (176,142) -- (176,182) -- (136,182) -- cycle ;
\draw  [fill={rgb, 255:red, 74; green, 74; blue, 74 }  ,fill opacity=0.65 ] (136,262) -- (176,262) -- (176,302) -- (136,302) -- cycle ;
\draw  [fill={rgb, 255:red, 155; green, 155; blue, 155 }  ,fill opacity=0.5 ] (96,262) -- (136,262) -- (136,302) -- (96,302) -- cycle ;
\draw  [fill={rgb, 255:red, 74; green, 74; blue, 74 }  ,fill opacity=0.65 ] (56,262) -- (96,262) -- (96,302) -- (56,302) -- cycle ;
\draw  [fill={rgb, 255:red, 74; green, 74; blue, 74 }  ,fill opacity=0.65 ] (176,222) -- (216,222) -- (216,262) -- (176,262) -- cycle ;
\draw  [fill={rgb, 255:red, 155; green, 155; blue, 155 }  ,fill opacity=0.5 ] (136,222) -- (176,222) -- (176,262) -- (136,262) -- cycle ;
\draw  [fill={rgb, 255:red, 74; green, 74; blue, 74 }  ,fill opacity=0.65 ] (486,142) -- (646,142) -- (646,302) -- (486,302) -- cycle ;
\draw  [draw opacity=0] (272,142) -- (433,142) -- (433,302.5) -- (272,302.5) -- cycle ; \draw   (272,142) -- (272,302.5)(292,142) -- (292,302.5)(312,142) -- (312,302.5)(332,142) -- (332,302.5)(352,142) -- (352,302.5)(372,142) -- (372,302.5)(392,142) -- (392,302.5)(412,142) -- (412,302.5)(432,142) -- (432,302.5) ; \draw   (272,142) -- (433,142)(272,162) -- (433,162)(272,182) -- (433,182)(272,202) -- (433,202)(272,222) -- (433,222)(272,242) -- (433,242)(272,262) -- (433,262)(272,282) -- (433,282)(272,302) -- (433,302) ; \draw    ;
\draw  [fill={rgb, 255:red, 155; green, 155; blue, 155 }  ,fill opacity=0.5 ] (272,142) -- (352,142) -- (352,222) -- (272,222) -- cycle ;
\draw  [fill={rgb, 255:red, 74; green, 74; blue, 74 }  ,fill opacity=0.65 ] (352,142) -- (432,142) -- (432,222) -- (352,222) -- cycle ;
\draw  [fill={rgb, 255:red, 74; green, 74; blue, 74 }  ,fill opacity=0.65 ] (272,222) -- (352,222) -- (352,302) -- (272,302) -- cycle ;
\draw  [fill={rgb, 255:red, 155; green, 155; blue, 155 }  ,fill opacity=0.5 ] (352,222) -- (432,222) -- (432,302) -- (352,302) -- cycle ;

\draw (146.5,324) node [anchor=north west][inner sep=0.75pt]    {$2^{1}$};
\draw (185.5,324) node [anchor=north west][inner sep=0.75pt]    {$2^{1}$};
\draw (166,328) node [anchor=north west][inner sep=0.75pt]   [align=left] {$\times$};
\draw (71,324) node [anchor=north west][inner sep=0.75pt]   [align=left] {\large{block size:}};
\draw (287,324) node [anchor=north west][inner sep=0.75pt]   [align=left] {\large{block size:}};
\draw (382,328) node [anchor=north west][inner sep=0.75pt]   [align=left] {$\times$};
\draw (401.5,324) node [anchor=north west][inner sep=0.75pt]    {$2^{2}$};
\draw (362.5,324) node [anchor=north west][inner sep=0.75pt]    {$2^{2}$};
\draw (500,324) node [anchor=north west][inner sep=0.75pt]   [align=left] {\large{block size:}};
\draw (596,328) node [anchor=north west][inner sep=0.75pt]   [align=left] {$\times$};
\draw (615,324) node [anchor=north west][inner sep=0.75pt]    {$2^{3}$};
\draw (576,324) node [anchor=north west][inner sep=0.75pt]    {$2^{3}$};

\end{tikzpicture}}
    
    \vspace{0.2cm}  
    
    \caption{Block sizes for an image of size $8 \times 8$.}
    \label{fig:blocks}
    
    \vspace{-0.2cm}
    
\end{figure}
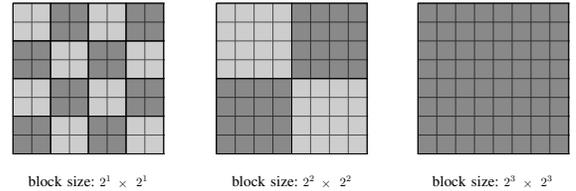

\paragraph{Rotation and image encryption} \label{ph:rotation_and_image_encryption}
As mentioned earlier, the generated keystream $s$ is the same size as the original image. Before image encryption, $s$, $I$, and $l$ are divided into square blocks from sizes $\{2^1 \times 2^1, ..., 2^n \times 2^n\}$ - where $n = \min(\lfloor M/2 \rfloor, \lfloor N/2 \rfloor)$. If the image size and the generated sequence $s$ cannot be fully divided into the excepted blocks, we retain the remaining pixels and do not rotate them. After partitioning the image into square blocks, the summation of each individual square in $s$ is calculated. Afterwards, modulus 4 is applied to the summation value to determine the rotation angle for each block. The mapping of each resulting value and the corresponding rotation angle is listed in the following: $\{{0: 0^{\circ}, 1: 90^{\circ}, 2: 180^{\circ}, 3: 270^{\circ}}\}$. The rotation process is expressed in Eq. \ref{eq: roration} and is demonstrated in Fig. \ref{fig:blocks}:
\begin{equation}
 \resizebox{.98\hsize}{!}{$
    \sum_{\scriptscriptstyle{i = C_12^n}}^{\scriptscriptstyle{(C_1+1)2^n}}\sum_{\scriptscriptstyle{j = C_22^n}}^{\scriptscriptstyle{(C_2+1)2^n}}s(i, j) \mod \; 4 \; = \\
    \begin{cases}
    0, \; \text{no rotation on $I$ and $l$} \\
    1, \; \text{rotate $I$ and $l$ 90 degree}\\
    2, \; \text{rotate $I$ and $l$ 180 degree} \\
    3, \; \text{rotate $I$ and $l$ 270 degree} 
    \end{cases}
    $}
    \label{eq: roration}
\end{equation}Where $C_1 \in [0, \floor{N/2^n}], C_1 \in \mathbf{Z}$, $C_2 \in [0, \floor{M/2^n}], C_2 \in \mathbf{Z}$, and $n \in \mathbf{Z}$. 

After the rotation, we encrypt the rotated original image using the stream cipher $s$ through exclusive-or (XOR) operation, and then we obtain the encrypted image $I_e$:

\begin{equation}
    I_e(i, j) = s(i, j) \oplus I(i, j)
    \label{eq: image_encryption}
\end{equation}

\paragraph{Location map embedding}
The next step is to embed the rotated location map directly into the LSBs of $I_e$. Among all those altered LSBs, there is a 50\% chance of a pixel's LSB being altered since the rotated location map and LSBs only contain 0s and 1s. As a result, the theoretically expected PSNR value of a reconstructed image using EMR-RDHEI is approximated as follows:
\begin{equation}
    \text{PSNR} \approx 10 \times \log _{10}\left(\frac{255^2}{1/2}\right) \approx 51.1411 \; dB
\end{equation}

\subsubsection{Data hiding phase}
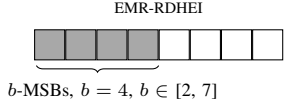
\begin{figure}[ht]
    \centering
    \begin{tikzpicture}[
    [dashed_box_node/.style={rectangle, draw=black, dashed, ultra thin},
    no_box_node/.style={rectangle, text width=1.2cm, align=center, minimum
                        width=1cm, minimum height=0.2cm, fill=white, font=\notsotiny},
    hard_box_node/.style={rectangle, draw=black, ultra thin, minimum width=1.7cm, 
                          minimum height=0.7cm, text width=1.7cm, align=center},
    X/.style = {rectangle, draw,
                minimum width=0.5cm, minimum height=0.5cm, text width=0.5cm,
                outer sep=0cm, inner sep=0cm, align=center, font=\scriptsize},
    Y/.style = {X, text width=0.4cm, minimum width=0.4cm, minimum height=0.4cm},
    red_box_node/.style = {X, draw=red, thick}]
    
    \scriptsize
    
    \definecolor{color1}{rgb}{0.66, 0.66, 0.66}
    
    
    \newcommand{\centerShift}{2cm}
    
    
    \node [no_box_node] at (-\centerShift, 0.5cm) {EMR-RDHEI};
    \matrix [matrix of nodes, nodes in empty cells, nodes={Y, anchor=center},
    column 1/.append style={nodes={fill=color1}}, 
    column 2/.append style={nodes={fill=color1}}, 
    column 3/.append style={nodes={fill=color1}},
    column 4/.append style={nodes={fill=color1}},
    ] at (-\centerShift, 0)(emr_bits)
    	{
             & & & & & & &\\
    	};
    	
    \draw[decoration={brace,mirror,raise=5pt},decorate] (-3.625cm,-0.1cm) -- node[below=9pt, xshift=0.2cm] {$b$-MSBs, $b = 4$, $b \in$ [2, 7]} (-2cm,-0.1cm);

    	
    
\end{tikzpicture}
    \caption{Example of EMR-RDHEI data hiding allocation.}
    \label{fig:emr_data_hiding}
    \vspace{-0.2cm}
\end{figure}

In the data hiding phase, the data hider shall not access the original image content $I$ nor the secret key $K_1$. First, the data hider extracts the rotated location map using the LSBs from the encrypted image $I_e$. After that, the data hider uses the secret key $K_2$ to encrypt the secret information. Then, the data hider's encrypted secret message is embedded using the pixels marked as 0 (redundant) by the location map into $I_e$. Thus, those redundant pixels' $b$-MSBs are replaced (shown in Fig. \ref{fig:emr_data_hiding}) and used for embedding data bits. Note that the pixels marked with a label of 1 in the location map cannot be changed. Since they remain unchanged in this process, they can be used as indicators during the reconstruction of the altered pixels' $b$-MSBs in the decoding phase.

\subsubsection{Decoding phase}
In the decoding phase, the receiver obtains the marked encrypted image $I_e^{'}$ and scans it in the same sliding fashion as previously mentioned. The receiver reconstructs the original image and/or extracts the hidden message depending on the available keys. Since our EMR-RDHEI method is separable, data extraction and image reconstruction can be performed independently. Using this method, we are able to recover the error-free secret message using the secret key $K_2$ and/or reconstruct the image $I'$ with low loss while using secret key $K_1$. There are three scenarios based on the key(s) that the receiver owns, which are discussed in the following.

\paragraph{The receiver has only the image encryption key $K_1$} 
\label{ph:image_derotation}
In this situation, the receiver can only recover the original image without deciphering the embedded secret message. With EMR-RDHEI, the receiver first extracts the location map $l$ from the LSBs of the marked encrypted image. Afterwards, the receiver decrypts the image using an XOR operation. Since the original image and location map are rotated before encryption, the next step is to perform the same sequence of rotations but in the opposite order. Similar to the encoding phase, we partition the 2D sequence $s$, marked decrypted image, and location map $l$ into certain blocks (depending on the block sizes used in the encoding phase). 
Unlike in Section \ref{ph:rotation_and_image_encryption}, we use reverse order block partitioning and apply the opposite rotation manner to de-rotate the previously rotated images. After calculating the summation of the blocks and applying modulus 4 to them, the rotation angles to undo the initial rotations are determined. The mapping of each remainder and inverse rotation angles is defined as: $\{{0: 0^{\circ}, 1: 270^{\circ}, 2: 180^{\circ}, 3: 90^{\circ}}\}$. The de-rotation process equation is the reverse of Eq. \ref{eq: roration2} and is expressed as follows:

\begin{equation}
 \resizebox{.98\hsize}{!}{$
    \sum_{\scriptscriptstyle{i = C_12^n}}^{\scriptscriptstyle{(C_1+1)2^n}}\sum_{\scriptscriptstyle{j = C_22^n}}^{\scriptscriptstyle{(C_2+1)2^n}}s(i, j) \mod \; 4 \; = \\
    \begin{cases}
    0, \; \text{no rotation on $I$ and $l$} \\
    1, \; \text{rotate $I$ and $l$ 270 degree}\\
    2, \; \text{rotate $I$ and $l$ 180 degree}\\
    3, \; \text{rotate $I$ and $l$ 90 degree} 
    \end{cases}
    $}
    \label{eq: roration2}
\end{equation} Where $C_1 \in [0, \floor{N/2^n}], C_1 \in \mathbf{Z}$, $C_2 \in [0, \floor{M/2^n}], C_2 \in \mathbf{Z}$, and $n \in \mathbf{Z}$.

After rotating the marked decrypted image and location map $l$, the original image $I$ is reconstructed by using $l$. The reconstruction process starts with initializing the $\omega$ vector as it copies the image's first column pixels' $b$-MSBs. This is done because the first column of the image is always labeled as 1 (non-redundant) in $l$. $\omega$ keeps track of the last non-redundant $b$-MSBs. The 1s (non-redundant) in the location map act as markers that help restore the $b$-MSBs of its following columns' pixels labeled as 0 (redundant) in $l$. By gliding $\omega$ along the width of $l$ and updating its value, we can quickly reconstruct $I$. This process is similar to the $l$ location map generation; however, this time, instead of identifying and annotating redundant pixels' $b$-MSBs, we reconstruct the redundant pixels' $b$-MSBs through the use of non-redundant pixels' $b$-MSBs. Following the procedures mentioned above, the original image can be reconstructed.

\paragraph{The receiver has only the data hiding key $K_2$} 
For this case, the receiver can only decipher the embedded secret message without reconstructing the original image. In order to decipher the embedded secret message, the receiver restores the location map $l$ from the LSBs of $I_e^{'}$. There is no need to undo the rotations (Eq. \ref{eq: roration2}) in this scenario because the secret message is embedded after the rotations. Then, the receiver concatenates all the $b$-MSBs of the pixels labeled with a 0 in $l$ and decrypts the concatenated bits using the secret key $K_2$ to obtain the plain message. With the proposed EMR-RDHEI method, we achieve error-free plain message recovery.

\paragraph{The receiver has both keys, $K_1$ and $K_2$} 
In the last scenario, the original image and plain message can be recovered by following the procedures described above.


\subsection{LMR-RDHEI method}

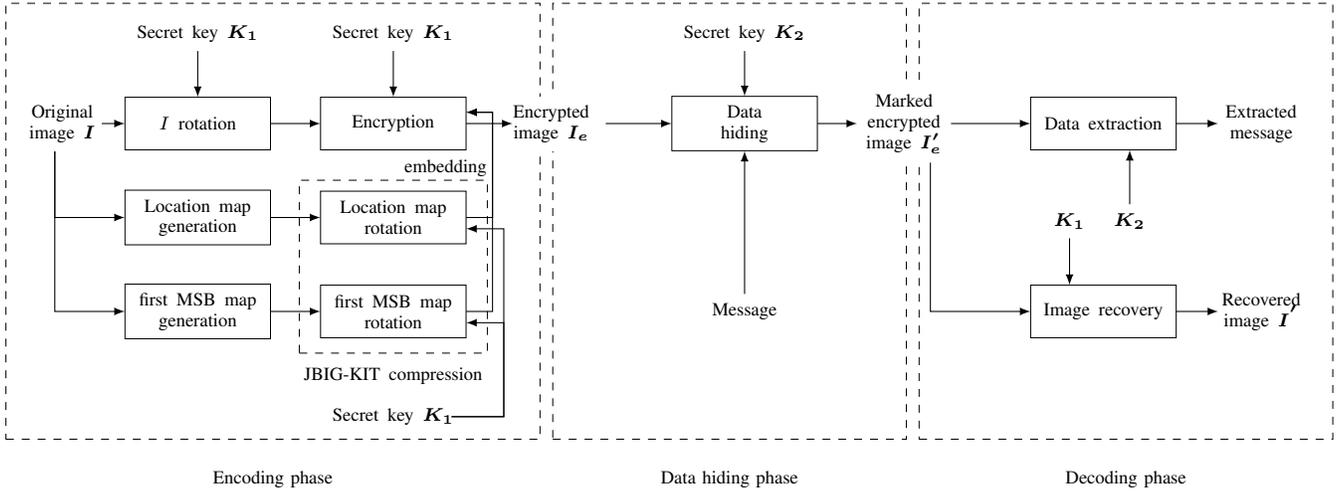
\begin{figure*}[ht]
\centering

    \begin{tikzpicture}[
    dashed_box_node/.style={rectangle, draw=black, dashed, ultra thin},
    no_box_node/.style={rectangle, text width=1.2cm, align=center, minimum width=1cm, minimum height=0.2cm, fill=white},
    hard_box_node/.style={rectangle, draw=black, ultra thin, minimum width=1.7cm, minimum height=0.7cm, text width=1.7cm, align=center}]
    
    \newcommand\x{0} 
    \newcommand\y{1.3}
    \newcommand\boxsize{5.8cm}
    \newcommand\gridboxYshift{0.0cm}
    \newcommand\dhc{6.07} 
    \newcommand\rpc{11.34} 
    
	\node [dashed_box_node, minimum width=7.1cm, minimum height=\boxsize] at (\x, \gridboxYshift) (content_owner_d_box) {};
	\node [dashed_box_node, minimum width=4.70cm, minimum height=\boxsize] at (\x+\dhc, \gridboxYshift) (data_hider_d_box) {};
	\node [dashed_box_node, minimum width=5.50cm, minimum height=\boxsize] at (\x+\rpc, \gridboxYshift) (recipient_d_box) {};
	
	\node [below=\boxsize-1.3cm] at (\x,\y) {\scriptsize Encoding phase};
	\node [below=\boxsize-1.3cm] at (\x+\dhc,\y) {\scriptsize Data hiding phase};
	\node [below=\boxsize-1.3cm] at (\x+\rpc,\y) {\scriptsize Decoding phase};
	
	\node [no_box_node, text width=1cm] at (\x-2.8, \y) (ORIGINAL_IMAGE) {\scriptsize Original \\[-1.5mm] image \bm{$I$}};
	\node [hard_box_node, text width=1.7cm] at (\x-1, \y) (I_ROTATION) {\scriptsize $I$ rotation};
	\node [no_box_node, text width=2cm, minimum width=2cm] at (\x-1, \y+1.2) (SECRET_KEY_1_0) {\scriptsize Secret key \bm{$K_1$}};
	\node [hard_box_node, text width=1.7cm] at (\x-1, \y-1.25) (LOCATION_MAP_GENERATION) {\scriptsize \scriptsize Location map \\[-1.5mm] generation};
	
	\node [hard_box_node, text width=1.7cm] at (\x-1, \y-2.5) (MSB_MAP_GENERATION) {\scriptsize \scriptsize first MSB map \\[-1.5mm] generation};
	
	\node [no_box_node, text width=2cm, minimum width=2cm] at (\x+1.6, \y+1.2) (SECRET_KEY_1_1) {\scriptsize Secret key \bm{$K_1$}};
	\node [hard_box_node] at (\x+1.6, \y) (encryption_h_box) {\scriptsize Encryption};
	
	\node [no_box_node, text width=1cm, minimum width=1cm, minimum width=0.1cm, minimum height=0.1cm,] at (\x+2.25, \y-0.59) (embedding) {\scriptsize embedding};
	
	\node [hard_box_node] at (\x+1.6, \y-1.25) (LOCATION_MAP_ROTATION) {\scriptsize Location map \\[-1.5mm] rotation};
	
	\node [hard_box_node] at (\x+1.6, \y-2.5) (MSB_MAP_ROTATION) {\scriptsize first MSB map \\[-1.5mm] rotation};
	
	\node [no_box_node, text width=2cm, minimum width=2cm] at (\x+1.6, \y-3.9) (SECRET_KEY_1_2) {\scriptsize Secret key \bm{$K_1$}};
	
	\node [dashed_box_node, minimum height=2.3cm, minimum width=2.5cm] at (\x+1.6, \y-0.65cm) (jbig_kit_compression) {};
    \node [no_box_node, text width=3cm, minimum width=1cm, minimum width=0.1cm, minimum height=0.1cm] at (\x+1.6, \y-2.1cm) {\scriptsize JBIG-KIT compression};

	\node [no_box_node, minimum width = 0.8cm, text width = 1.2cm, minimum height = 0.4] at (\x+3.7, \y) (encrypted_image_n_box) {\scriptsize Encrypted \\[-1.5mm] image \bm{$I_e$}};
	
	\node [no_box_node, text width=2cm, minimum width=2cm] at (\x+\dhc+0.2, \y+1.2) (secret_key_w_n_box) {\scriptsize Secret key \bm{$K_2$}};
	\node [hard_box_node] at (\x+\dhc+0.2, \y) (data_hiding_h_box) {\scriptsize Data \\[-1.5mm] hiding};
	\node [no_box_node, text width=0.9cm] at (\x+\dhc+0.2, \y-2.5) (message_n_box) {\scriptsize Message};

	\node [no_box_node, text width=1.3cm, minimum width=0.1cm] at (\x+\rpc-2.95, \y) (marked_encrypted_image_n_box) {\scriptsize Marked \\[0mm] encrypted \\[-1.5mm] image \bm{$I_{e}'$}};
	
	\node [hard_box_node] at (\x+\rpc-0.3, \y-2.5) (image_recovery_h_box) {\scriptsize Image recovery};
	\node [hard_box_node] at (\x+\rpc-0.3, \y) (data_extraction_h_box) {\scriptsize Data extraction};
    \node [no_box_node, text width=0.5cm, minimum width=0.5cm] at (\rpc-0.75, \x) (secret_key_1_2_n_box) {\scriptsize \bm{$K_1$}};
    \node [no_box_node, text width=0.5cm, minimum width=0.5cm] at (\rpc+0.05, \x) (secret_key_2_2_n_box) {\scriptsize \bm{$K_2$}};
    \node [no_box_node] at (\x+\rpc+1.8, \y-2.5) (reconstructed_image_n_box) {\scriptsize Recovered \\[-1.5mm] image \bm{$I^{'}$}};
    \node [no_box_node] at (\x+\rpc+1.8, \y) (extracted_message_n_box) {\scriptsize Extracted \\[-1.5mm] message};

    \draw[arrows={-latex}] ([xshift=-0.1cm]ORIGINAL_IMAGE.east) -- (I_ROTATION.west);
    \draw[arrows={-latex}] ([xshift=-0.1cm]ORIGINAL_IMAGE.south) |- (LOCATION_MAP_GENERATION.west);
    \draw[arrows={-latex}] ([xshift=-0.1cm]ORIGINAL_IMAGE.south) |- (MSB_MAP_GENERATION.west);
    \draw[arrows={-latex}] (I_ROTATION.east) -- (encryption_h_box.west);
    \draw[arrows={-latex}] (LOCATION_MAP_GENERATION.east) -- (LOCATION_MAP_ROTATION.west);
    \draw[arrows={-latex}] (MSB_MAP_GENERATION.east) -- (MSB_MAP_ROTATION.west);
    \draw[arrows={-latex}] (SECRET_KEY_1_0.south) -- (I_ROTATION.north);
    \draw[arrows={-latex}] (SECRET_KEY_1_1.south) -- (encryption_h_box.north);
    \draw[arrows={-latex}] ([xshift=-0.35cm]SECRET_KEY_1_2.east) -| ([xshift=0.5cm, yshift=-0.15cm]MSB_MAP_ROTATION.east) -- ([yshift=-0.15cm]MSB_MAP_ROTATION.east);
    \draw[arrows={-latex}] ([xshift=-0.35cm]SECRET_KEY_1_2.east) -| ([xshift=0.5cm, yshift=-0.15cm]LOCATION_MAP_ROTATION.east) -- ([yshift=-0.15cm]LOCATION_MAP_ROTATION.east);
    
    \draw[arrows={-latex}] (LOCATION_MAP_ROTATION.east) -| ([xshift=0.35cm, yshift=0.15cm]encryption_h_box.east) -- ([yshift=0.15cm]encryption_h_box.east);
    \draw[arrows={-latex}] (MSB_MAP_ROTATION.east) -| ([xshift=0.35cm, yshift = 0.15cm]encryption_h_box.east) -- ([yshift=0.15cm]encryption_h_box.east);
    
    \draw[arrows={-latex}] (encryption_h_box.east) -- ([xshift=0.21cm]encrypted_image_n_box.west);
    \draw[arrows={-latex}] (encrypted_image_n_box.east) -- (data_hiding_h_box.west);
    
    \draw[arrows={-latex}] (secret_key_w_n_box.south) -- (data_hiding_h_box.north);
    \draw[arrows={-latex}] (message_n_box.north) -- (data_hiding_h_box.south);
    \draw[arrows={-latex}] (data_hiding_h_box.east) -- ([xshift=0.15cm]marked_encrypted_image_n_box.west);
    
    \draw[arrows={-latex}] ([xshift=-0.15cm]marked_encrypted_image_n_box.east) -- (data_extraction_h_box.west);
    \draw[arrows={-latex}] ([xshift=0.35cm]marked_encrypted_image_n_box.south) |- (image_recovery_h_box.west);
    \draw[arrows={-latex}] (secret_key_1_2_n_box.south) -- ([xshift=-0.45cm]image_recovery_h_box.north);
    \draw[arrows={-latex}] (secret_key_2_2_n_box.north) -- ([xshift=0.35cm]data_extraction_h_box.south);
    \draw[arrows={-latex}] (data_extraction_h_box.east) -- ([xshift=0.15cm]extracted_message_n_box.west);
    \draw[arrows={-latex}] (image_recovery_h_box.east) -- ([xshift=0.15cm]reconstructed_image_n_box.west);

\end{tikzpicture}
    
    \vspace{0.2cm}  
    
    \caption{General Schema of LMR-RDHEI Method.}
    \label{fig:General_LMR}
    
    \vspace{-0.2cm}
    
\end{figure*}

The general schema of our proposed LMR-RDHEI method is demonstrated in Fig. \ref{fig:General_LMR}. The encoding phase in LMR-RDHEI includes the following procedures: (1) location map and first MSB map generation, (2) location map and first MSB map compression, (3) original image, location map, and MSB map rotation, (4) original image encryption, and (5) compressed maps embedding. In both the first MSB map and location map's compression processes, we employ the JBIG-KIT lossless image compression library \citep{JBIG-KIT}. The original image, first MSB map, and location map use secret key $K_1$ to rotate. The encryption of the original image also uses secret key $K_1$ to encrypt. 

In the data hiding phase, we apply the multi-MSB replacement technique to embed a secret message into the encrypted image's redundant pixels. Since the first MSB bit-plane stores the bits of the compressed rotated first MSB map and compressed rotated location map, we do not insert any other bit information into the first MSB bit-plane. Thus, compared with the EMR-RDHEI method, the overall embedding rate slightly decreases. At the final decoding phase, the hidden message is extracted from the marked encrypted image without any errors, and the original image is recovered losslessly. Similar to the EMR-RDHEI method, the LMR-RDHEI method is separable, meaning that the data extraction and image recovery processes can be done independently. The LMR-RDHEI method is also processed in a vectorized sliding window manner.

\subsubsection{Encoding phase} \label{sssc:lmr_encoding_phase}

\paragraph{Maps generation, rotation, and compression}
First, the most optimal location map $l$ is generated with a process similar\footnote{$b \in [2, 7]$ in EMR-RDHEI, and $b \in [2, 8]$ in LMR-RDHEI.} to the procedure we described in Section \ref{ph:optimal_location_map_generation}. Then, the first MSB map, which is extracted from the first MSB bit-plane, can be expressed as the following:
\begin{equation}
    \eta(i, j) = (I(i, j) \; \land \; 128)\mod \;\; 127
    \label{eq: MSB_map}
\end{equation}

After generating the most optimal location map $l$ and the first MSB map $\eta$, we rotate them (along with the original image) according to the secret keystream $s$ generated by $K_1$. Note that the secret key generation process is the same one illustrated in Section \ref{ph:secret_key_generation}. The overall rotation procedure is also similar to the steps we specified in Section \ref{ph:rotation_and_image_encryption}. For the LMR-RDHEI method, we take the JBIG-KIT compression efficiency and future data embedding rate into consideration when choosing block sizes. The rotation procedure can negatively impact the JBIG-KIT compression efficiency because the shuffling lowers the correlation between new neighboring pixels. To further explain, the following example is provided. For images of size 512 $\times$ 512, instead of using the block sizes of \{$2^1 \times 2^1$, $2^2 \times 2^2$, ... , $2^9 \times 2^9$\}, block sizes of \{$2^4 \times 2^4$, $2^5 \times 2^5$, ... , $2^9 \times 2^9$\} are adopted primarily to gain a better compression rate by JBIG-KIT. By using these larger block sizes, a comparably good compression rate can be achieved, a high message embedding rate is maintained, and the images remain secure. For more details, experiments regarding different block sizes will be demonstrated in Section \ref{ssc:detailed_examples}. In this rotation process, if the compressed maps' size exceeds the size of the first MSB bit-plane (512 $\times$ 512 bits), we generate the next most optimal location map $l$ until both maps can be sufficiently compressed to fit in the first MSB bit-plane collectively. We choose to alter the location map because the compressed first MSB map is fixed.

\paragraph{Maps embedding and image encryption}
In contrast to the previously mentioned EMR-RDHEI method, LMR-RDHEI can realize a lossless recovery of the original image. EMR-RDHEI stores the location map into LSBs of an encrypted image, and it will cause a slight bit-level loss in the restored image. LMR-RDHEI concatenates the compressed first MSB and location maps then inserts them back into the first MSB bit-plane of an encrypted image to prevent any potential bit losses. After the rotated image is encrypted using the generated secret key $K_1$ with Eq. \ref{eq: image_encryption}, we embed both maps into the encrypted image.

\subsubsection{Data hiding phase}
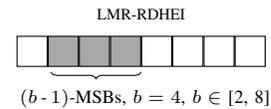
\begin{figure}[ht]
    \centering
    \begin{tikzpicture}[
    [dashed_box_node/.style={rectangle, draw=black, dashed, ultra thin},
    no_box_node/.style={rectangle, text width=1.2cm, align=center, minimum
                        width=1cm, minimum height=0.2cm, fill=white, font=\notsotiny},
    hard_box_node/.style={rectangle, draw=black, ultra thin, minimum width=1.7cm, 
                          minimum height=0.7cm, text width=1.7cm, align=center},
    X/.style = {rectangle, draw,
                minimum width=0.5cm, minimum height=0.5cm, text width=0.5cm,
                outer sep=0cm, inner sep=0cm, align=center, font=\scriptsize},
    Y/.style = {X, text width=0.4cm, minimum width=0.4cm, minimum height=0.4cm},
    red_box_node/.style = {X, draw=red, thick}]
    
    \scriptsize
    
    \definecolor{color1}{rgb}{0.66, 0.66, 0.66}
    
    
    \newcommand{\centerShift}{2cm}
    
    
    	

    \node [no_box_node] at (\centerShift, 0.5cm) {LMR-RDHEI};
    \matrix [matrix of nodes, nodes in empty cells, nodes={Y, anchor=center},
    column 2/.append style={nodes={fill=color1}}, 
    column 3/.append style={nodes={fill=color1}},
    column 4/.append style={nodes={fill=color1}}
    ] at (\centerShift, 0)(amr_bits)
    	{
            & & & & & & &\\
    	};
    	
    \draw[decoration={brace,mirror,raise=5pt},decorate] (0.8cm,-0.1cm) -- node[below=9pt, xshift=0.65cm] {$(b\,$-$\,1)$-MSBs, $b = 4$, $b \in$ [2, 8]} (2cm,-0.1cm);
    
\end{tikzpicture}
    \caption{Example of LMR-RDHEI data hiding allocation.}
    \label{fig:lmr_data_hiding}
    \vspace{-0.2cm}
\end{figure}

The first MSB bit-plane cannot be modified because it holds essential information of the compressed first MSB map and location map; therefore, it is preserved for the decoding phase. The data hider extracts the compressed location map from the first MSB bit-plane and then decompresses it using the JBIG-KIT. The data hider then hides the secret bits into the pixels labeled with a 0 in the location map. Like EMR-RDHEI, the pixels labeled with a 1 cannot be used for embedding any secret information. These pixels serve as indicators that are essential for recovering the original ($b$-1) MSB bits in those altered pixels. As for pixels marked with a label of 0 in the location map, ($b$-1) out of $b$ bits can be replaced and used for embedding a secret message. For example, if $b$ = 4, only 3 of 4 bits (excluding the first bit) in a pixel can hide secret information (shown in Fig. \ref{fig:lmr_data_hiding}).

\subsubsection{Decoding phase}
Similar to the aforementioned EMR-RDHEI method, LMR-RDHEI is separable --- the data extraction and image recovery processes can be performed individually. A receiver can extract the hidden message and/or reconstruct the original image depending on the available secret key(s). Unlike EMR-RDHEI, not only can LMR-RDHEI recover error-free secret message with secret key $K_2$, but it can also realize a lossless image recovery. There are three scenarios in the decoding phase, which are presented as follows.

\paragraph{The receiver has only the image encryption key $K_1$}
For recovering the original image losslessly only, the receiver can first extract the compressed location map $l$ and the compressed first MSB map $\eta$ from the first MSB bit-plane of the marked encrypted image. After decompressing both maps with JBIG-KIT, the receiver uses the secret key $K_1$ to decrypt the image with an XOR operation and obtain the decrypted marked image. Since the two maps and the original image are rotated before encryption, the next step is to de-rotate both maps and the decrypted marked image with the secret key $K_1$. The process for de-rotating these maps and images is performed in the same manner as the one described in Section \ref{ph:image_derotation} with the only minor difference of adding the de-rotation process to the first MSB map in LMR-RDHEI. After obtaining the de-rotated marked decrypted image, $l$, and $\eta$, the decompressed de-rotated $\eta$ is used to recover the original first MSB bit-plane. In the end, the original image $I$ is restored losslessly by decompressing the de-rotated $l$ and later recovering the ($b$-1) MSBs with its guidance. The ($b$-1) MSBs recovery is similar to EMR-RDHEI's $b$ MSBs recovery process described in the latter part of Section \ref{ph:image_derotation}. The only difference is that the location map $l$ is used to restore the ($b$-1) MSBs instead of the $b$ MSBs.

\paragraph{The receiver has only the data hiding key $K_2$}
For recovering the plain message only, the receiver first extracts the compressed location map from the first MSB bit-plane. Afterwards, JBIG-KIT is applied to decompress the location map. Similar to EMR-RDHEI, there is no de-rotation (Eq. \ref{eq: roration2}) in this procedure. The next step is to concatenate all the ($b$-1)-MSBs of pixels labeled as 0 in the corresponding location map. Finally, the concatenated bits are decrypted using the data hiding key $K_2$. Thus, the plain message is obtained. Note that our LMR-RDHEI method can realize error-free plain message recovery.

\paragraph{The receiver has both keys, $K_1$ and $K_2$} 
For the last case, the receiver can follow the steps described above for recovering the original image and the hidden message, resulting in a lossless image and a error-free plain message.

\section{Experimental Results} \label{sc:results}
In this section, we present experimental results to demonstrate and support our proposed methods. In Section \ref{ssc:detailed_examples}, a detailed example is provided: the grey-level image Lena of size $512 \times 512$ using EMR-RDHEI and LMR-RDHEI. After that, 10,000 images with various textures from the BOWS-2 database \citep{BOWS-2} are used to measure the general efficiency of our methods. The security analyses of our methods are presented in Section \ref{ssc:security_analysis}. Finally, a comparison of our methods with many other state-of-the-art schemas is shown in Section \ref{ssc:comparison_with_other_approaches}. In Sections \ref{ssc:security_analysis} and \ref{ph:performance_comparison}, PSNR and SSIM are used to measure the similarity between target samples.

\begin{figure}[ht]
    \centering
    
    \scriptsize

    {\input{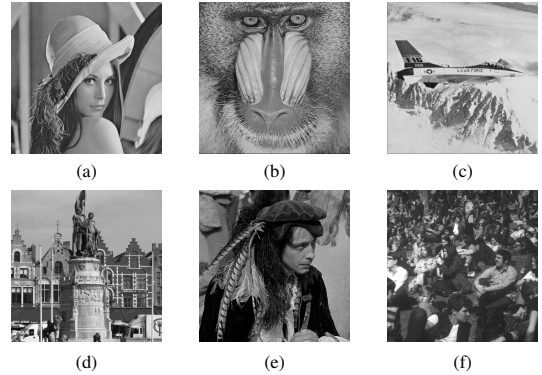}}
    
    \caption{Standard grey-level images of size $512 \times 512$: a) Lena; b) Baboon; c) Airplane; d) Knight; e) Man; f) Crowd.}
    \label{fig:analysis}
    
\end{figure}

Throughout our experiments, the samples used are standard grey-level images of size $512 \times 512$, which are listed in Fig. \ref{fig:analysis}. Additionally, for testing these images, the chosen block sizes for EMR-RDHEI are \{$2^1 \times 2^1$, $2^2 \times 2^2$, ..., $2^9 \times 2^9$\}. As mentioned in Section \ref{ph:optimal_location_map_generation}, for images of size $512 \times 512$, the block sizes of \{$2^4 \times 2^4$, $2^5 \times 2^5$, ... , $2^9 \times 2^9$\} are chosen for LMR-RDHEI to ensure JBIG-KIT's compression efficiency, maintain a high embedding rate, and secure the overall information.

\subsection{Detailed examples} \label{ssc:detailed_examples}
\begin{figure}[ht]
    \centering

    \input{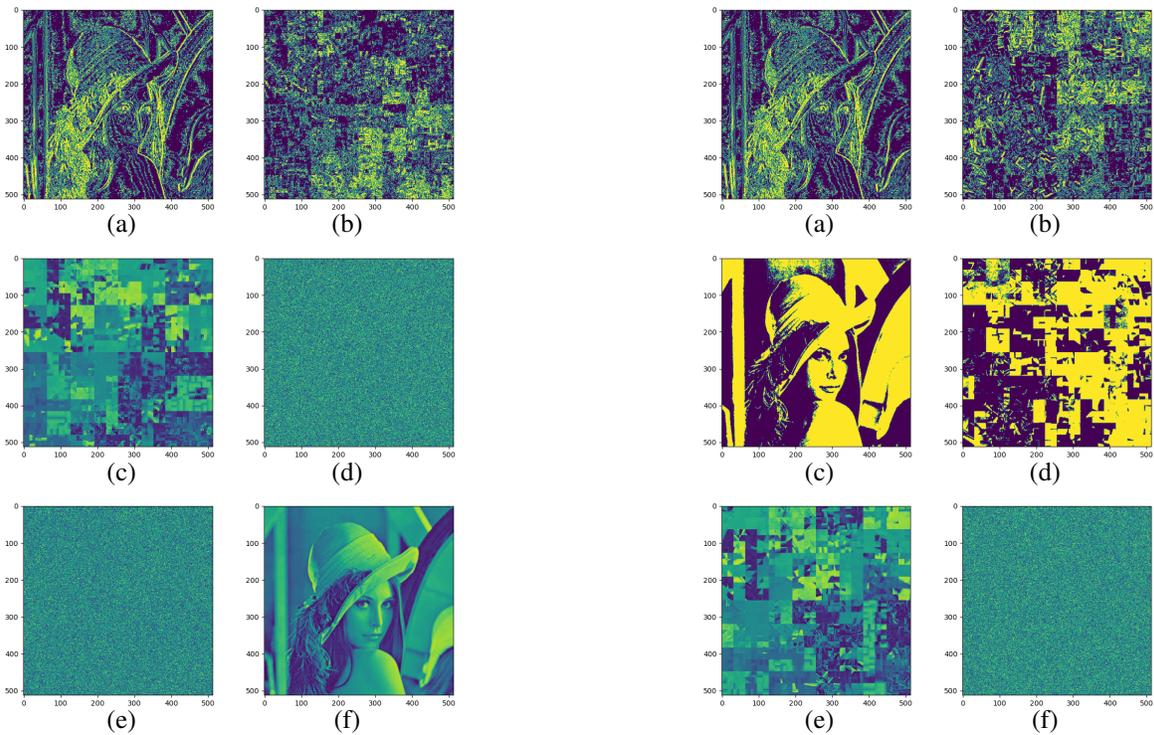}
    
    \caption{With our EMR-RDHEI method, the embedding rate was 2.6566 bpp. a) The most optimal generated location map $l$, when $b = 4$; b) The location map after rotation; c) The original image Lena $I$ after rotation; d) Encrypted image Lena $I_e$; e) Marked encrypted image Lena $I_e^{'}$; f) Reconstructed image $I^{'}$, PSNR = 51.1356 dB, SSIM = 0.9928.}
    
    \label{fig:EMR_Lena}
    
\end{figure}

\begin{figure}[ht]
    \centering

    \input{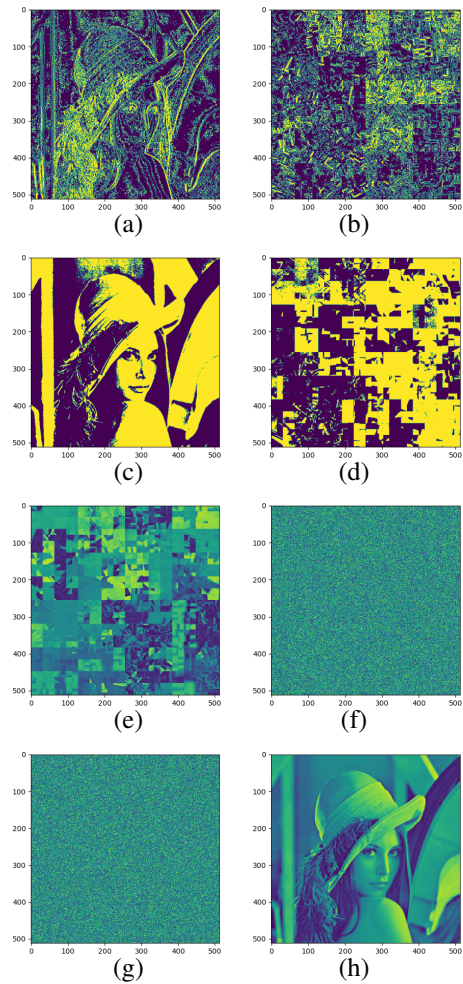}
    
    \caption{With our LMR-RDHEI method, the embedding rate was 1.9925 bpp. a) The most optimal generated location map $l$, when $b = 4$; b) The location map after rotation; c) The extracted first MSB map before rotation; d) The first MSB map after rotation; e) The original image Lena $I$ after rotation; f) Encrypted image Lena $I_e$; g) Marked encrypted image Lena $I_e^{'}$; h) Reconstructed image $I^{'}$, PSNR $\rightarrow$ +$\infty$, SSIM = 1.}
    
    \label{fig:LMR_Lena}
    
\end{figure}

We thoroughly demonstrate the performance of EMR-RDHEI and LMR-RDHEI by applying them to the grey-level image Lena of size $512 \times 512$. Results are shown in Fig. \ref{fig:EMR_Lena} and Fig. \ref{fig:LMR_Lena}. Additionally, to measure the general efficiency of both EMR-RDHEI and LMR-RDHEI, our methods are tested using 10,000 grey-level images with various textures from BOWS-2 database \citep{BOWS-2}. In our experiments, we compute the minimum, maximum, and average values of DER, PSNR, and SSIM. Furthermore, as mentioned earlier in Section \ref{sssc:lmr_encoding_phase}, we test LMR-RDHEI with different block sizes. The results are summarized in Table \ref{Tab:EMR_results} and Table \ref{Tab:LMR_block_size}. In Table \ref{Tab:LMR_block_size}, DER indicates the average embedding rate. When testing LMR-RDHEI, good cases mean both the first MSB map and location map are successfully compressed to fit in the first MSB bit-plane. In contrast, bad cases imply that both maps cannot be compressed sufficiently; therefore, the image cannot be used to hide any secret message using LMR-RDHEI.
\vspace{0.25cm}

\renewcommand{\arraystretch}{1.9}
\begin{table}[h]

    \centering
    \scriptsize
    
    \begin{center}
        \begin{tabular}{ |c|c|c|c|c| }
            \cline{3-5}
            \multicolumn{2}{c|}{}
                                                    & DER (bpp)  & PSNR (dB) & SSIM\\
            \hline
            \multicolumn{2}{|c|}{maximum}    
                                                    & \emrMaxDER{} & \emrMaxPSNR{} & \emrMaxSSIM{} \\
            \hline
            \multicolumn{2}{|c|}{average}    
                                                    & \emrAverageDER{} & \emrAveragePSNR{} & \emrAverageSSIM{} \\
            \hline
            \multicolumn{2}{|c|}{minimum}     
                                                    & \emrMinDER{} & \emrMinPSNR{} & \emrMinSSIM{} \\
            \hline
        \end{tabular}
    \end{center}
    
    \vspace{-0.40cm}
    
    \caption{Results base on EMR-RDHEI method.}
    \label{Tab:EMR_results}

\end{table}

\renewcommand{\arraystretch}{1}
\renewcommand{\arraystretch}{1.5}
\begin{table}[H]
    \centering
    \footnotesize
    \begin{tabular}{ |c|c|c|c|c| } 
     \hline
     block sizes & DER (bpp) & good cases & bad cases \\
     \hline
     \{$2^4 \times 2^4, ..., 2^9 \times 2^9$\} & \lmrAverageDER{} & 99.99\% & 0.01\% \\ 
     \hline
     \{$2^3 \times 2^3, ..., 2^9 \times 2^9$\} & 2.5118 & 99.98\% & 0.02\% \\ 
     \hline
     \{$2^2 \times 2^2, ..., 2^9 \times 2^9$\} & 2.4737 & 99.93\% & 0.07\% \\ 
     \hline
     \{$2^1 \times 2^1, ..., 2^9 \times 2^9$\} & 2.4402 & 99.86\% & 0.14\% \\ 
     \hline
    \end{tabular}
    \vspace{-0.20cm}
    
    \caption{Testing different block sizes on the BOWS-2.}
    \label{Tab:LMR_block_size}
\end{table}

\begin{figure}[H]
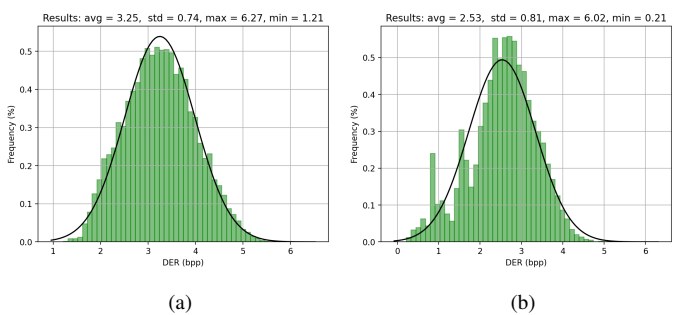

    \centering
    \captionsetup[sub]{font=small}
    
    \newcommand{\bppGraphSize}{4.4cm}
    \newcommand{\bppLeftShift}{-0.2cm}

    \subfloat[]
        {\label{fig:DER_a}\hspace{\bppLeftShift}\input{graphs/EMR_DER}}\subfloat[]
        {\label{fig:DER_b}\hspace{\bppLeftShift}\input{graphs/LMR_DER}}
    
    
    \caption{DER results of the total 10,000 samples from the BOWS-2 database. DER analysis for (a) EMR-RDHEI and (b) LMR-RDHEI methods.}
    \label{fig: LMR_EMR_DER}
    
    \vspace{-0.2cm}
    
\end{figure}

To demonstrate the dynamic range of embedding rates using both methods, we sample 10,000 images from the BOWS-2 dataset and plot the DER histograms in Fig. \ref{fig: LMR_EMR_DER}\subref{fig:DER_a} and Fig. \ref{fig: LMR_EMR_DER}\subref{fig:DER_b} for EMR-RDHEI and LMR-RDHEI. The average embedding rates were \emrAverageDER{} bpp and \lmrAverageDER{} bpp, respectively.

\renewcommand{\arraystretch}{2}
\begin{table*}[ht]
    \centering
    \footnotesize
    
    \begin{center}
    \renewcommand{\arraystretch}{1.3}
        \begin{tabular}{ |c|c|c|p{1.7cm}|p{2.1cm}|p{2.0cm}|p{2.0cm}|p{1.7cm}| }
            \hline
            \multicolumn{1}{|c|}{Images} &
            \multicolumn{2}{c|}{Methods} &
            \multicolumn{1}{c|}{Entropy} & \multicolumn{1}{c|}{$\chi^2$ test} & 
            \multicolumn{1}{c|}{NPCR (\%)} & 
            \multicolumn{1}{c|}{UACI (\%)} & 
            \multicolumn{1}{c|}{PSNR (dB)}\\
            \hline
            \multirow[c]{6}*{Lena} &
            \multirow[c]{3}*{EMR-RDHEI/LMR-RDHEI} & \makecell{$I$} & \makecell{7.4456} & \makecell{340.6470} & \makecell{/} & \makecell{/} & \makecell{/}\\
            \cline{3-8}
            & & \makecell{$I_e$} & \makecell{7.9994/7.9994} & \makecell{15.0197/14.8271} & \makecell{99.6025/99.5853} & \makecell{28.7115/28.7276} & \makecell{9.2217/9.2041}\\ 
            \cline{3-8}
            & & \makecell{$I_e^{'}$} & \makecell{7.8583/7.9993} & \makecell{169.1861/15.4492} & \makecell{99.6250/99.6044} & \makecell{28.6808/28.6693} & \makecell{9.2173/9.2261}\\ 
            \cline{2-8}
            & \multirow[c]{3}*{\citet{Asad2020}} & \makecell{$I$} & \makecell{7.4456} & \makecell{340.6470} & \makecell{/} & \makecell{/} & \makecell{/}\\
            \cline{3-8}
            & & \makecell{$I_e$} & \makecell{7.9994} & \makecell{14.6164} & \makecell{99.6223} & \makecell{28.6375} & \makecell{9.2247}\\ 
            \cline{3-8}
            & & \makecell{$I_e^{'}$} & \makecell{7.9993} & \makecell{15.7695} & \makecell{99.6262} & \makecell{28.6367} & \makecell{9.2265}\\
            \thickhline
            \multirow[c]{6}*{Baboon} &
            \multirow[c]{3}*{EMR-RDHEI/LMR-RDHEI} & \makecell{$I$} & \makecell{7.3579} & \makecell{396.0794} & \makecell{/} & \makecell{/} & \makecell{/}\\
            \cline{3-8}
            & & \makecell{$I_e$} & \makecell{7.9992/7.9993} & \makecell{16.8684/16.2893} & \makecell{99.6181/99.6216} & \makecell{27.8875/27.9002} & \makecell{9.5132/9.5074}\\ 
            \cline{3-8}
            & & \makecell{$I_e^{'}$} & \makecell{7.8588/7.9994} & \makecell{100.6018/14.2764} & \makecell{99.6204/99.6162} & \makecell{28.0922/27.8928} & \makecell{9.4575/9.5106}\\ 
            \cline{2-8}
            & \multirow[c]{3}*{\citet{Asad2020}} & \makecell{$I$} & \makecell{7.3579} & \makecell{396.0794} & \makecell{/} & \makecell{/} & \makecell{/}\\
            \cline{3-8}
            & & \makecell{$I_e$} & \makecell{7.9992} & \makecell{16.1595} & \makecell{99.6155} & \makecell{27.8212} & \makecell{9.5270}\\ 
            \cline{3-8}
            & & \makecell{$I_e^{'}$} & \makecell{7.9993} & \makecell{15.7457} & \makecell{99.6143} & \makecell{27.8899} & \makecell{9.5194}\\
            \thickhline
            \multirow[c]{6}*{Airplane} &
            \multirow[c]{3}*{EMR-RDHEI/LMR-RDHEI} & \makecell{$I$} & \makecell{6.6776} & \makecell{826.2035} & \makecell{/} & \makecell{/} & \makecell{/}\\
            \cline{3-8}
            & & \makecell{$I_e$} & \makecell{7.9993/7.9993} & \makecell{16.3967/16.4838} & \makecell{99.5983/99.5991} & \makecell{32.3403/32.3001} & \makecell{8.0536/8.0641}\\ 
            \cline{3-8}
            & & \makecell{$I_e^{'}$} & \makecell{7.9351/7.9993} & \makecell{115.1221/15.9512} & \makecell{99.6048/99.6201} & \makecell{32.3150/32.3569} & \makecell{8.0593/8.0504}\\ 
            \cline{2-8}
            & \multirow[c]{3}*{\citet{Asad2020}} & \makecell{$I$} & \makecell{6.6776} & \makecell{826.2035} & \makecell{/} & \makecell{/} & \makecell{/}\\
            \cline{3-8}
            & & \makecell{$I_e$} & \makecell{7.9993} & \makecell{16.4092} & \makecell{99.6101} & \makecell{32.2851} & \makecell{8.0674}\\ 
            \cline{3-8}
            & & \makecell{$I_e^{'}$} & \makecell{7.9993} & \makecell{16.1689} & \makecell{99.6132} & \makecell{32.3564} & \makecell{8.0516}\\
            \thickhline
            \multirow[c]{6}*{Knight} &
            \multirow[c]{3}*{EMR-RDHEI/LMR-RDHEI} & \makecell{$I$} & \makecell{7.3227} & \makecell{661.7007} & \makecell{/} & \makecell{/} & \makecell{/}\\
            \cline{3-8}
            & & \makecell{$I_e$} & \makecell{7.9992/7.9992} & \makecell{16.1288/16.6209} & \makecell{99.6021/99.6212} & \makecell{30.1702/30.1601} & \makecell{8.7040/8.7064}\\ 
            \cline{3-8}
            & & \makecell{$I_e^{'}$} & \makecell{7.9574/7.9993} & \makecell{76.7231/15.4173} & \makecell{99.6216/99.5953} & \makecell{30.1915/30.1196} & \makecell{8.7002/8.7175}\\ 
            \cline{2-8}
            & \multirow[c]{3}*{\makecell{\citet{Puteaux2018} \\ (CPE-HCRDH/EPE-HCRDH)}} & \makecell{$I$} & \makecell{7.3227} & \makecell{668.628} & \makecell{/} & \makecell{/} & \makecell{/}\\
            \cline{3-8}
            & & \makecell{$I_e$} & \makecell{7.9994/7.9994} & \makecell{14.8342/14.8806} & \makecell{99.6143/99.6136} & \makecell{30.1338/30.1344} & \makecell{8.7081/8.7081}\\ 
            \cline{3-8}
            & & \makecell{$I_e^{'}$} & \makecell{7.9994/7.9994} & \makecell{15.1188/14.8299} & \makecell{99.6082/99.6059} & \makecell{30.1521/30.1569} & \makecell{8.7069/8.7039}\\
            \hline
        \end{tabular}
    \end{center}
    
    \vspace{-0.20cm}
    
    \caption{Security evaluation on classic images with our proposed methods.}
    \label{Tab:analysis}

\end{table*}
\renewcommand{\arraystretch}{1}

\subsection{Security analysis}
\label{ssc:security_analysis}
In this subsection, we use some well-known statistical metrics, including Shannon entropy, $\chi^2$ test, number of changing pixel rate (NPCR), and unified averaged changed intensity (UACI) tests to analyze the security level of our proposed methods. Formal definitions for each metric are given below, followed by testing results.

\paragraph{Shannon entropy}
This measures the probability distribution of different pixel values in an image. A higher Shannon entropy value indicates that the pixel values in an image are distributed more uniformly within an allowable range. The Shannon entropy, $H(I)$, of an image $I$ is defined as follows:
\begin{equation}
    H(I) = -\sum_{i=0}^{255}P(\alpha_i)\log _{2}(P(\alpha_i))
\end{equation}
$P(\alpha_i)$ represents the probability of a pixel value $\alpha_i$ ($0 \leq \alpha_i \leq 255$) in each grey level.

\paragraph{$\chi^2$ test}
It is an indicator of divergence in a grey-level image $I$ from its theoretical counterpart in which all pixels occur with an equal probability of $1/256$. $\chi^2$ can be computed as follows:
\begin{equation}
    \chi^2 = 256 \cdot (h \times w) \sum_{i=0}^{255}\left(P(\alpha_i)-\frac{1}{256}\right)^2
\end{equation}

\paragraph{NPCR and UACI analysis}
These two tests are typical quantities used to evaluate the strength of an image encryption algorithm against differential attacks. Conventionally, a higher NPCR/UACI score is interpreted as a stronger resistance to differential attacks \citep{Wu2011}.
\begin{equation}
    NPCR = \frac{1}{h \times w}\sum_{i=0}^{h-1}\sum_{j=0}^{w-1}\sigma(i,j) \times 100\%,
\end{equation}
where:
\begin{equation}
    \sigma(i,j) = 
    \begin{cases}
    1, & \text{if} \; I(i,j) = I'(i,j)\\
    0, & \text{otherwise}
    \end{cases}
\end{equation}
and
\begin{equation}
    UACI = \frac{1}{h \times w} \sum_{i=0}^{h-1}\sum_{j=0}^{w-1} \frac{|I(i,j) - I'(i,j)|}{255} \times 100\%,
\end{equation} 
where $I(i, j)$ represents the pixels in the original grey-level image and $I'(i, j)$ represents the pixels in the reconstructed image.

In Table \ref{Tab:analysis}, $I$, $I_e$, and $I_e^{'}$ represent the original image, encrypted image, and marked encrypted image, respectively. For the statistical security analysis, our results show that the sensitive information is undetectable in the encrypted image and marked encrypted image using our proposed methods. As shown in Table \ref{Tab:analysis}, the similarity (in terms of PSNR) between an original image and an encrypted image/marked encrypted image is very low (approximately 9 dB). Furthermore, if part of the secret message is modified by an attacker, as the message is being encrypted with secret key $K_2$, they cannot be decrypted and exploited for validation in the later decoding phase. If an attacker attempts to modify $b$-MSBs in the pixels labeled with a 1 in the location map, significant noise will be introduced into the reconstructed image. Note that we also need to de-rotate the original image and the location map in both of our methods to perform image reconstruction. Thus, the original image content cannot be obtained, and these additional processes further secure the confidential information. Moreover, without the encryption key $K_1$, the marked encrypted image cannot be decrypted, de-rotated, and recovered.

\subsection{Comparison with related approaches} \label{ssc:comparison_with_other_approaches}
More comparisons of our proposed EMR-RDHEI and LMR-RDHEI methods with many other state-of-the-art approaches are demonstrated in this subsection. These include DER comparisons and performance comparisons. The test samples used in the DER and performance comparisons were previously shown in Fig. \ref{fig:analysis}.
\begin{figure*}[ht]
    \captionsetup[subfigure]{justification=centering}
    \centering
    
    \notsotiny
    \newcommand{\graphHeight}{3.5cm}
    \newcommand{\graphWidth}{4.5cm}
    \newcommand{\leftShift}{0cm}
    \hspace{-1.2cm}
    
    \begin{adjustbox}{minipage=\linewidth,scale=0.85}
    \hspace{-1.2cm}
    \subfloat[Test image: Lena.]{
        \label{fig:DC_a}
        \centering
        \definecolor{color1}{RGB}{157, 230, 244}
\definecolor{color2}{RGB}{140, 209, 235}
\definecolor{color3}{RGB}{122, 188, 226}
\definecolor{color4}{RGB}{105, 167, 217}
\definecolor{color5}{RGB}{87, 146, 208}
\definecolor{color6}{RGB}{70, 125, 199}
\definecolor{color7}{RGB}{52, 104, 190}
\definecolor{color8}{RGB}{35, 83, 181}
\definecolor{color9}{RGB}{17, 62, 172}
\definecolor{color10}{RGB}{0, 41, 163}
\definecolor{color11}{RGB}{0, 30, 155}


\hspace{-\leftShift}\begin{tikzpicture}[x={(0.01,0)}]
    
    \newcommand{\graphXScale}{\graphWidth / 2.5}
    \newcommand{\graphYScale}{\graphHeight / 10}
    \newcommand{\graphXmax}{3.3}
    
    \foreach  \l/\x/\c[count=\y] in 
    {
    \textbf{EMR-RDHEI}/2.657/color11, 
    \textbf{LMR-RDHEI}/1.993/color10,
    \citet{Chen2019}/1.972/color9,
    \citet{Puteaux2020}/1.810/color7,
    Puteaux and Puech (CPE) \citep{Puteaux2018}/1.000/color6,
    Puteaux and Puech (EPE) \citep{Puteaux2018}/0.964/color5,
    \citet{XCao2016}/0.780/color4,
    \citet{Asad2020}/0.748/color3,
    \citet{Liu2017}/0.686/color2,
    \citet{Puteaux2017}/0.500/color1}
    {
        \node[left] at (0,\y*\graphYScale) {\l};
        \fill[\c] (0,\y*\graphYScale-.4*\graphYScale) rectangle (\x*\graphXScale,\y*\graphYScale+.4*\graphYScale);
    }
    
    \draw (0,0) -- (\graphXmax*\graphXScale,0);
    \draw (0,0) -- (0,\graphHeight+1*\graphYScale);
    \draw (0,\graphHeight+1*\graphYScale) -- (\graphXmax*\graphXScale, \graphHeight+1*\graphYScale);
    \draw (\graphXmax*\graphXScale, 0) -- (\graphXmax*\graphXScale, \graphHeight+1*\graphYScale);
    
    \foreach \x in {0, 1, 2, ..., \graphXmax}{
        \draw (\x*\graphXScale,.125) -- (\x*\graphXScale,0) node[below] {\x};
    }
\end{tikzpicture}
    }\subfloat[Test database: BOWS-2.]{
        \label{fig:DC_b}
        \centering

\definecolor{color1}{RGB}{157, 230, 244}
\definecolor{color2}{RGB}{140, 209, 235}
\definecolor{color3}{RGB}{122, 188, 226}
\definecolor{color4}{RGB}{105, 167, 217}
\definecolor{color5}{RGB}{87, 146, 208}
\definecolor{color6}{RGB}{70, 125, 199}
\definecolor{color7}{RGB}{52, 104, 190}
\definecolor{color8}{RGB}{35, 83, 181}
\definecolor{color9}{RGB}{17, 62, 172}
\definecolor{color10}{RGB}{0, 41, 163}
\definecolor{color11}{RGB}{0, 30, 155}


\hspace{\leftShift}\begin{tikzpicture}[x={(.01,0)}]
    
    \newcommand{\graphXScale}{\graphWidth / 2.45}
    \newcommand{\graphYScale}{\graphHeight / 7}
    \newcommand{\graphXmax}{3.3}
    
    \foreach  \l/\x/\c[count=\y] in 
    {
    \textbf{LMR-RDHEI}/2.533/color11, 
    \citet{Puteaux2020}/2.459/color9,
    \citet{Chen2019}/2.245/color8,
    \citet{SYi2018}/1.881/color7,
    \citet{epe2018}/1.836/color5,
    \citet{YPuyang2018}/1.346/color3,
    Puteaux and Puech (EPE) \citep{Puteaux2018}/0.968/color1}
    {
        \node[left] at (0,\y*\graphYScale) {\l};
        \fill[\c] (0,\y*\graphYScale-.4*\graphYScale) rectangle (\x*\graphXScale,\y*\graphYScale+.4*\graphYScale);
    }

    \draw (0,0) -- (\graphXmax*\graphXScale,0);
    \draw (0,0) -- (0,\graphHeight+1*\graphYScale);
    \draw (0,\graphHeight+1*\graphYScale) -- (\graphXmax*\graphXScale, \graphHeight+1*\graphYScale);
    \draw (\graphXmax*\graphXScale, 0) -- (\graphXmax*\graphXScale, \graphHeight+1*\graphYScale);
    
    \foreach \x in {0, 1, 2, ..., \graphXmax}{
        \draw (\x*\graphXScale,.125) -- (\x*\graphXScale,0) node[below] {\x};
    }
\end{tikzpicture}
    }

    \caption{Comparisons of (maximum/average) DERs with other state-of-the-art methods on image Lena and BOWS-2 database.}
    \label{fig:DERComparison}
    \end{adjustbox}
    
    \vspace{-0.2cm}
    
\end{figure*}
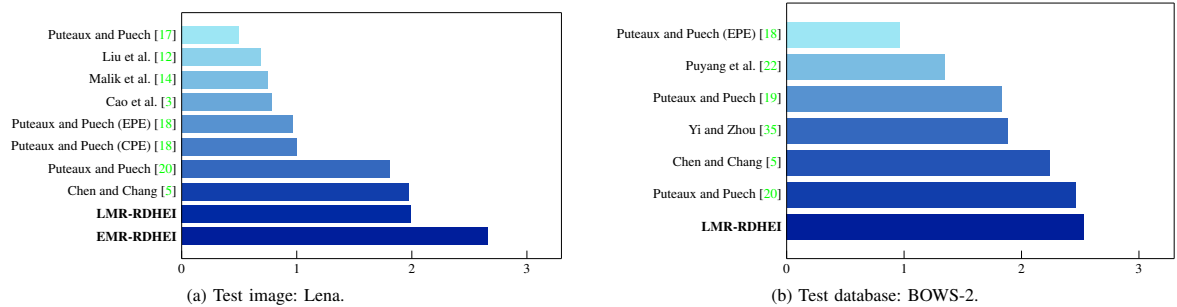
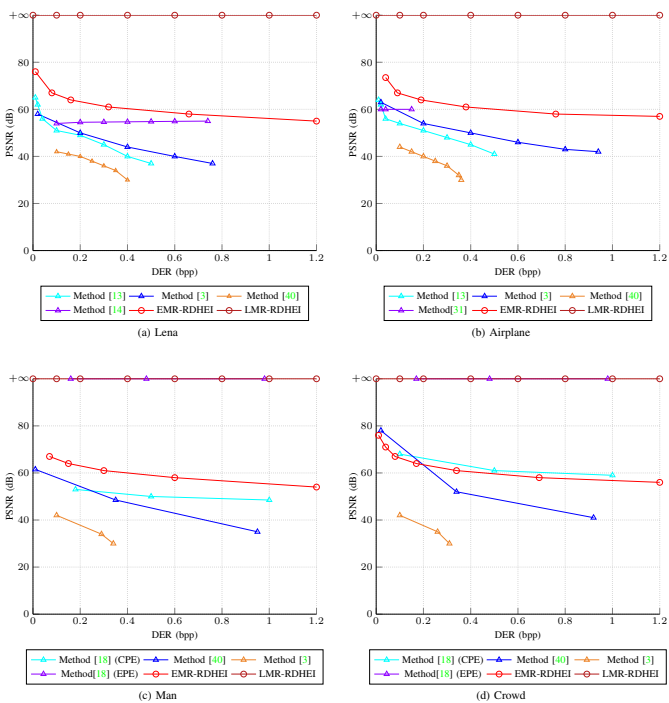
\begin{figure}[ht]
    \centering
    \scriptsize
    \bigskip
    \begin{adjustbox}{minipage=\linewidth,scale=0.55}
    \hspace{-4.3cm} 
    \subfloat[Lena]{
        \centering
        \definecolor{aqua}{rgb}{0.0, 1.0, 1.0}
\definecolor{blue}{rgb}{0.0, 0.0, 1.0}
\definecolor{cadmiumorange}{rgb}{0.93, 0.53, 0.18}
\definecolor{carnelian}{rgb}{0.7, 0.11, 0.11}
\definecolor{electricviolet}{rgb}{0.56, 0.0, 1.0}
\begin{tikzpicture}
\scriptsize
    \begin{axis}[
        xlabel = DER (bpp),
        ylabel = PSNR (dB),
        xlabel near ticks,
        ylabel shift = -9 pt,
        axis x line*=bottom,
        axis y line*=left,
        xmin=0, xmax=1.2,
        ymin=0, ymax=100,
        xtick={0,0.2,0.4,0.6,0.8,1,1.2},
        ytick={0,20,40,60,80},
        extra y ticks={100},
        extra y tick labels={$+\infty$},
        ymajorgrids=true,
        xmajorgrids=true,
        grid style=densely dotted,
        legend style={at={(0.5,-0.15)},anchor=north},
        legend columns=3]
    
    \addplot[ 
        color=aqua,
        mark=triangle,
        ]
        coordinates {
        (0.01,65) (0.02,62) (0.04,56) (0.1, 51) (0.2,49) (0.3,45) (0.4,40) (0.5,37)};
        
    \addplot[ 
        color=blue,
        mark=triangle,
        ]
        coordinates {
        (0.02,58) (0.2,50) (0.4,44) (0.6,40) (0.76,37)};

    \addplot[ 
        color=cadmiumorange,
        mark=triangle,
        mark size=1.5pt,
        ]
        coordinates {
        (0.1,42) (0.15,41) (0.2,40) (0.25,38) (0.3,36) (0.35,34) (0.4,30)};
        
    \addplot[ 
        color=electricviolet,
        mark=triangle,
        ]
        coordinates {
        (0.1,54) (0.2,54.5) (0.3,54.6) (0.4,54.7) (0.5,54.8) (0.6,54.9) (0.74,55)};
    
    \addplot[ 
        color=red,
        mark=o,
        ]
        coordinates {(0.01, 76) (0.08, 67)
        (0.16, 64) (0.32, 61) (0.66, 58) (1.2, 55)};
        
    \addplot[ 
        color=carnelian,
        mark=o,
        ]
        coordinates {
        (0, 100) (0.1, 100) (0.2,100) (0.4,100) (0.6,100) (0.8,100) (1.0,100) (1.2,100)};

    \legend{Method \citep{KMa2013}, Method \citep{XCao2016}, Method \citep {XZhang2016}, Method \citep{Asad2020}, EMR-RDHEI, LMR-RDHEI};
    

    \end{axis}
\end{tikzpicture}
    }\hspace{-0.3cm}
    \subfloat[Airplane]{
        \centering
        \definecolor{aqua}{rgb}{0.0, 1.0, 1.0}
\definecolor{blue}{rgb}{0.0, 0.0, 1.0}
\definecolor{cadmiumorange}{rgb}{0.93, 0.53, 0.18}
\definecolor{carnelian}{rgb}{0.7, 0.11, 0.11}
\definecolor{electricviolet}{rgb}{0.56, 0.0, 1.0}
\begin{tikzpicture}
    \scriptsize
    \begin{axis}[
        xlabel = DER (bpp),
        ylabel = PSNR (dB),
        xlabel near ticks,
        ylabel shift = -9 pt,
        xmin=0, xmax=1.2,
        ymin=0, ymax=100,
        axis x line*=bottom,
        axis y line*=left,
        xtick={0,0.2,0.4,0.6,0.8,1,1.2},
        ytick={0,20,40,60,80},
        extra y ticks={100},
        extra y tick labels={$+\infty$},
        legend pos=north west,
        ymajorgrids=true,
        xmajorgrids=true,
        grid style=densely dotted,
        legend style={at={(0.5,-0.15)},anchor=north},
        legend columns=3
        ]
        
    \addplot[ 
        color=aqua,
        mark=triangle,
        ]
        coordinates {
        (0.01,64) (0.02,62) (0.04,56) (0.1, 54) (0.2,51) (0.3,48) (0.4,45) (0.5,41)};
        
     \addplot[ 
        color=blue,
        mark=triangle,
        ]
        coordinates {
        (0.02,63) (0.2,54) (0.4,50) (0.6,46) (0.8,43) (0.94,42)};
        
    \addplot[ 
        color=cadmiumorange,
        mark=triangle,
        ]
        coordinates {
        (0.1,44) (0.15,42) (0.2,40) (0.25,38) (0.3,36) (0.35,32) (0.36,30)};
        
     \addplot[ 
        color=electricviolet,
        mark=triangle,
        ]
        coordinates {
        (0.02,60) (0.04,60) (0.15,60)};
    
    \addplot[ 
        color=red,
        mark=o,
        ]
        coordinates {(0.04, 73.5) (0.09, 67)
        (0.19, 64) (0.38, 61) (0.76, 58) (1.2, 57)};
        
    \addplot[ 
        color=carnelian,
        mark=o
        ]
        coordinates {
        (0, 100) (0.1, 100) (0.2,100) (0.4,100) (0.6,100) (0.8,100) (1.0,100) (1.2,100)};

    \legend{Method \citep{KMa2013}, Method \citep{XCao2016}, Method \citep{XZhang2016}, Method\citep{XWu2014}, EMR-RDHEI, LMR-RDHEI};

    \end{axis}
\end{tikzpicture}
    }
    
    \bigskip
    
    \hspace{-4.3cm} 
    \subfloat[Man]{
        \centering
        \definecolor{aqua}{rgb}{0.0, 1.0, 1.0}
\definecolor{blue}{rgb}{0.0, 0.0, 1.0}
\definecolor{cadmiumorange}{rgb}{0.93, 0.53, 0.18}
\definecolor{carnelian}{rgb}{0.7, 0.11, 0.11}
\definecolor{electricviolet}{rgb}{0.56, 0.0, 1.0}
\begin{tikzpicture}
    \scriptsize



    \begin{axis}[
        xlabel = DER (bpp),
        ylabel = PSNR (dB),
        xlabel near ticks,
        ylabel shift = -9 pt,
        xmin=0, xmax=1.2,
        ymin=0, ymax=100,
        axis x line*=bottom,
        axis y line*=left,
        xtick={0,0.2,0.4,0.6,0.8,1,1.2},
        ytick={0,20,40,60,80},
        extra y ticks={100},
        extra y tick labels={$+\infty$},
        legend pos=north west,
        ymajorgrids=true,
        xmajorgrids=true,
        grid style=densely dotted,
        legend style={at={(0.5,-0.15)},anchor=north},
        legend columns=3
        ]
        
     \addplot[ 
        color=aqua,
        mark=triangle,
        ]
        coordinates {
        (0.18,53) (0.5,50) (1,48.5)};
        
     \addplot[ 
        color=blue,
        mark=triangle,
        ]
        coordinates {
        (0.01,61.5) (0.35,48.5) (0.95,35)};
        
    \addplot[ 
        color=cadmiumorange,
        mark=triangle,
        ]
        coordinates {
        (0.1,42) (0.29,34) (0.34,30)};
        
    \addplot[ 
        color=electricviolet,
        mark=triangle,
        ]
        coordinates {
        (0.16,100) (0.48,100) (0.98, 100)};
    
    \addplot[ 
        color=red,
        mark=o,
        ]
        coordinates {(0.07, 67)
        (0.15, 64) (0.3, 61) (0.6, 58) (1.2, 54)};
        
    \addplot[ 
        color=carnelian,
        mark=o,
        ]
        coordinates {
        (0, 100) (0.1, 100) (0.2,100) (0.4,100) (0.6,100) (0.8,100) (1.0,100) (1.2,100)};

    \legend{Method \citep{Puteaux2018} (CPE), Method \citep{XZhang2016}, Method \citep{XCao2016}, Method\citep{Puteaux2018} (EPE), EMR-RDHEI, LMR-RDHEI};

    \end{axis}
\end{tikzpicture}
    }\hspace{-0.3cm}
    \subfloat[Crowd]{
        \centering
        \definecolor{aqua}{rgb}{0.0, 1.0, 1.0}
\definecolor{blue}{rgb}{0.0, 0.0, 1.0}
\definecolor{cadmiumorange}{rgb}{0.93, 0.53, 0.18}
\definecolor{carnelian}{rgb}{0.7, 0.11, 0.11}
\definecolor{electricviolet}{rgb}{0.56, 0.0, 1.0}
\begin{tikzpicture}

    \scriptsize
    \begin{axis}[
        xlabel = DER (bpp),
        ylabel = PSNR (dB),
        xlabel near ticks,
        ylabel shift = -9 pt,
        xmin=0, xmax=1.2,
        ymin=0, ymax=100,
        axis x line*=bottom,
        axis y line*=left,
        xtick={0,0.2,0.4,0.6,0.8,1,1.2},
        ytick={0,20,40,60,80},
        extra y ticks={100},
        extra y tick labels={$+\infty$},
        legend pos=north west,
        ymajorgrids=true,
        xmajorgrids=true,
        grid style=densely dotted,
        legend style={at={(0.5,-0.15)},anchor=north},
        legend columns=3
        ]
        
     \addplot[ 
        color=aqua,
        mark=triangle,
        ]
        coordinates {
        (0.1,68) (0.5,61) (1,59)};
        
     \addplot[ 
        color=blue,
        mark=triangle,
        ]
        coordinates {
        (0.02,78) (0.34,52) (0.92,41)};
        
    \addplot[ 
        color=cadmiumorange,
        mark=triangle,
        ]
        coordinates {
        (0.1,42) (0.26,35) (0.31,30)};
        
    \addplot[ 
        color=electricviolet,
        mark=triangle,
        ]
        coordinates {
        (0.17,100) (0.48,100) (0.98, 100)};
    
    \addplot[ 
        color=red,
        mark=o,
        ]
        coordinates {(0.01, 76) (0.04, 71) (0.08, 67)
        (0.17, 64) (0.34, 61) (0.69, 58) (1.2, 56)};
        
    \addplot[ 
        color=carnelian,
        mark=o,
        ]
        coordinates {
        (0, 100) (0.1, 100) (0.2,100) (0.4,100) (0.6,100) (0.8,100) (1.0,100) (1.2,100)};

    \legend{Method \citep{Puteaux2018} (CPE), Method \citep{XZhang2016}, Method \citep{XCao2016}, Method\citep{Puteaux2018} (EPE), EMR-RDHEI, LMR-RDHEI};

    \end{axis}
\end{tikzpicture}
    }
   
    \caption{Performance comparison with related works.}
    \label{fig:PerformanceComparison}
    \end{adjustbox}
    
    \vspace{-0.2cm}
    
\end{figure}

\subsubsection{DER comparison}
In order to demonstrate a straightforward embedding rate comparison between our proposed methods and other recent methods published by \citep{XCao2016, Puteaux2017, Liu2017, Puteaux2018, Asad2020, Chen2019, Puteaux2020, SYi2019}, we present the maximum DERs of the image Lena and the average DERs of 10,000 images from BOWS-2 image dataset in Fig. \ref{fig:DERComparison}.

The maximum DERs obtained by \citep{XCao2016, Puteaux2017, Liu2017, Puteaux2018, Asad2020, Chen2019, Puteaux2020} are shown in Fig. \ref{fig:DERComparison}\subref{fig:DC_a}. Most methods are not able to achieve a maximum embedding rate higher than 2 bpp. In the most recent research published by \citep{Asad2020, Chen2019, Puteaux2020}, their best DERs underperform our results: 2.6566 bpp and 1.9925 bpp for EMR-RDHEI and LMR-RDHEI, respectively. As seen in Fig. \ref{fig:DERComparison}\subref{fig:DC_a}, our data hiding rate of EMR-RDHEI is significantly higher than the recently published works by \citet{Puteaux2020, Chen2019, Puteaux2018}.

In Fig. \ref{fig:DERComparison}\subref{fig:DC_b}, we compare the average embedding rate of LMR-RDHEI with other lossless image recovery methods: \citep{Puteaux2018, YPuyang2018, epe2018, SYi2018, Chen2019, Puteaux2020}. The average DERs produced by \citep{Puteaux2018, YPuyang2018, epe2018, SYi2018} are all less than 2 bpp. In 2018, \citet{Puteaux2018} proposed their EPE-HCRDH approach, which yielded an average DER of 0.968 bpp. Later, \citet{YPuyang2018} employed a two-MSB prediction schema and obtained an average DER of 1.346 bpp. In comparison to recently published works by \citep{Chen2019, Puteaux2020}, the average DER result of \lmrAverageDER{} bpp from our LMR-RDHEI method is a substantial improvement compared with previously published methods.

\subsubsection{Performance comparison} \label{ph:performance_comparison}
In this section, we use the following standard grey-level images of size $512 \times 512$: Lena, Airplane, Man, and Crowd to perform the performance comparison. As shown in Fig. \ref{fig:PerformanceComparison}, the bit-level differences between the reconstructed images and original images are measured using PSNR and used to compare our results with current state-of-the-art methods \citep{KMa2013, XCao2016, XZhang2016, Asad2020, XWu2014, Puteaux2018}.

As can be seen from our results, our maximum embedding rates outperform previously published schemas. For images Lena, Airplane, Man, and Crowd, the maximum DER using EMR-RDHEI was 2.6566 bpp, 3.0574 bpp, 2.3917 bpp, 2.7504 bpp, respectively; and with LMR-RDHEI, we achieved the maximum DER of 1.9925 bpp, 2.4459 bpp, 1.5118 bpp, 2.2003 bpp, each to each. For LMR-RDHEI, a lossless image recovery can be realized in all testing samples. None of the aforementioned methods achieve the same high embedding rate while maintaining a high PSNR between restored and original images. When we test similar embedding rates (such as 0.5 bpp and 1.2 bpp) compared with other recent approaches, it is clear that the PSNRs obtained by our methods are close to or even higher than state-of-the-art works. In the performance comparisons of images Lena and Man, our LMR-RDHEI method outperforms the lossless recovery EPE-HCRDH method published by \citet{Puteaux2018} in terms of DER.

To summarize, both of our proposed methods simultaneously provide error-free message extraction and, to the best of our knowledge, have one of the highest performances in tests regarding the data embedding rate and image reconstruction quality found in the RDHEI literature. Our proposed methods, EMR-RDHEI and LMR-RDHEI, allow a better trade-off between data hiding rate and visual quality in restored images thanks to the multi-MSB replacement approach.

\section{Conclusion} \label{sc:conclusion}
In this paper, we propose two multi-MSB replacement-based RDHEI methods: EMR-RDHEI and LMR-RDHEI. Although the reconstructed images from the EMR-RDHEI method slightly differ from their original contents, in practice, the difference in their visual quality is indistinguishable since only the LSB of some pixels changed. On the other hand, our proposed LMR-RDHEI method is able to recover the original images perfectly without any loss. Through extensive experiments presented in Section \ref{sc:results}, we show that our algorithms outperform many other current state-of-the-art methods in both embedding rate and PSNR. Our novel strategy includes the multi-MSB replacement and the image block division/rotation, in contrast to previously suggested methods such as prediction error, histogram shifting, and other popular techniques. Overall, the multi-MSB replacement technique offers a cleaner design through vectorized bit operations, enabling faster and more efficient computation during image processing. A limitation to our work is that the LMR-RDHEI method has a very low probability ($\leq$ 0.14\%) of resulting in a bad case among testing samples, as shown in Table. \ref{Tab:LMR_block_size}. In Section \ref{ssc:detailed_examples}, we stated that bad cases occur mostly due to some images having high frequency noise and complex textures. It occurs when LMR-RDHEI does not sufficiently compress the assistant data to be stored in the first MSB plane. In the future, we aim to find better solutions to the challenge of achieving a reasonable compression rate with some highly-textured images. Given the high performance of our multi-MSB replacement-based technique and its principle of identifying and utilizing redundant bits to embed a secret message, further investigations on this method are recommended. For example, it is possible to apply a dynamic $b$ parameter to trace and use more redundant bits in an original image to achieve a better performance overall.

\section*{Acknowledgement}
The authors would like to thank the anonymous reviewers for their valuable time and advice.

\bibliography{main}

\vspace{-2cm}
\begin{IEEEbiography}[{\includegraphics[width=1in,height=1.25in,clip,keepaspectratio]{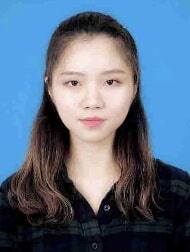}}]{Yike Zhang}
received an M.S. degree in Computer Engineering from the Engineering Department at St. Mary's University, San Antonio, Texas, USA in Spring 2021. She is currently pursuing her Ph.D. in Computer Science at Vanderbilt University. Her research interests include image processing, multimedia information security, digital forensics, and watermarking. Her work has focused on image processing and analysis in the encrypted domain.
\end{IEEEbiography}
\vspace{-2cm}
\begin{IEEEbiography}[{\includegraphics[width=1in,height=1.25in,clip,keepaspectratio]{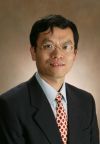}}]{Wenbin Luo} 
received the B.S. and M.S. degrees in electrical engineering from Fudan University, Shanghai, P.R.China. He received another M.S. degree in statistics and a Ph.D. degree in computer engineering from the University of New Mexico, Albuquerque, NM, USA. Also, he is an Oracle Certified Professional (OCP), a Senior Level Linux Professional (LPIC-3), and a Ubuntu Certified Professional (UCP). He is currently a professor of computer engineering at St. Mary's University of San Antonio, where he has been teaching since 2003. He has authored or co-authored over 40 research papers, a book, and two book chapters. His research interests include digital image processing, computer security, and data structures \& algorithms. He was the Publication Chair of IEEE SoSE08, SoSE09, \& SoSE13 and was a Program Co-Chair of IEEE SoSE15.
\end{IEEEbiography}

\end{document}